\documentclass[12pt]{JHEP3}
\usepackage{graphicx}
\def\ti    {\tilde}
\def\sq    {\ti q}
\def\gluino{\ti g}
\newcommand{\slush}{\!\!\!/}

\title{Handling jets + missing $E_T$ channel using inclusive $m_{T2}$ }
\author{Mihoko M. Nojiri\\ 
Theory Group, KEK and the Graduate University for Advanced Study
	(SOUKENDAI), Oho 1-1, Tukuba, Ibaraki, 305-0801, Japan,
\\
Institute for the Physics and Mathematics of the Universe (IPMU),
 University of Tokyo, 5-1-5 Kashiwa-noHa, Kashiwa, Chiba, 277-8568, Japan}
\author{Kazuki Sakurai\\ 
Department of Physics, Nagoya University 
\\Nagoya, Aichi, 464-8602, Japan}
\author{Yasuhiro Shimizu\\
 Tohoku University International Advanced Research and Education Organization, 
\\Institute for International Advanced Interdisciplinary Research, 
\\Sendai, Miyagi 980-8578, Japan 
} 
\author{Michihisa Takeuchi\\ 
Yukawa Institute for Theoretical Physics, Kyoto University,\\
 Kyoto 606-8502, Japan\\
Theory Group, KEK, 
Oho 1-1, Tukuba, Ibaraki, 305-0801, Japan
 }
\preprint{IPMU 08-0051, KEK TH-1267 ,TU-824, YITP-08-66}

\abstract{The ATLAS and CMS experiments at the Large Hadron Collider (LHC) 
may discover the squarks ($\tilde{q}$) and gluino ($\tilde{g}$) of the minimal supersymmetric 
standard model (MSSM) in the early stage of the experiments if their 
masses are lighter than 1.5 TeV.   
In this paper we propose  the sub-system $m_{T2}$ variable ($m^{\rm sub }_{T2}$),
which is sensitive to the gluino mass when $m_{\tilde{q}}>m_{\tilde{g}}$. 
Using  it with the inclusive $m_{T2}$ distribution proposed earlier, 
$\tilde{q}$ and $\tilde{g}$ masses can be determined simultaneously in the 
early stage of the experiments.  
Results of Monte Carlo simulations at
sample MSSM model points are presented both for signal and background. 
}
\keywords{Supersymmetry Phenomenology, Supersymmetric Standard Model }
\begin{document}
\setcounter{footnote}{0}
\section{Introduction}
Supersymmetry (SUSY) provides  an elegant  solution to 
the hierarchy problem in the Standard Model (SM)  Higgs sector
 \cite{Nilles:1983ge, Haber:1984rc,Martin:1997ns}.
It predicts a 
set of new particles containing spin 0 sfermions and spin 1/2 gauginos and 
higgsinos.   
If  R parity is conserved,  
the lightest supersymmetric particle (LSP), 
which is often the lightest neutralino,
is stable and a good dark matter candidate. 
The thermal relic density of the LSP can be consistent with the 
cold dark matter density of our Universe. 

The ATLAS and CMS experiments at the CERN Large Hadron Collider 
(LHC) may discover 
the SUSY particles in the early stage of data collection. 
The missing momentum carried by 
the stable LSP becomes an important signature of the sparticle 
production.  
Current studies show that an integrated luminosity of
around $1\,$fb$^{-1}$  is enough to find sparticle production  
if the squark and gluino  masses are below 
1.5 TeV  and the mass difference between the LSP and squark/gluino 
is large. 

We do not yet have many clues 
on the sparticle mass scale, although the current 
measurements of flavor changing neutral current (FCNC) give stringent constraints on 
the relation among  sfermion masses unless they 
are extremely heavy. 
Once we have seen signs of SUSY at the LHC,
we should use direct evidence
to determine the SUSY particle masses, 
from which we may determine the sparticle mass relations. 
Various methods have been developed 
for spaticle mass determination
from event kinematics. 
The invariant mass distributions 
of various exclusive channels are known to be very useful 
\cite{Hinchliffe:1996iu,Abdullin:1998pm,Atlas,
Bachacou:1999zb,Hinchliffe:1999zc,Allanach:2000kt}.
By combining the measured 
endpoints of the distributions of the relatively  
clean and long cascade decay channels involving 
charginos ($\tilde{\chi}^{\pm }_i$), neutralinos ($\tilde{\chi}^0_i$) and sleptons ($\tilde{l}$),
such as the opposite sign same flavor lepton signal 
arising from 
$\tilde{q}\rightarrow \tilde{\chi}^0_2 q\rightarrow 
\tilde{l}ql \rightarrow \tilde{\chi}^0_1 qll$, 
one can determine  not only the  masses of the squark and gluino, but also  
the  neutralino and slepton masses arising from their cascade decays. 
The exact relations 
 among  momenta of  visible particles from a cascade decay
 are also useful  \cite{Nojiri:2003tu, Kawagoe:2004rz,Nojiri:2007pq,Cheng:2008mg}. 
For some cases, 
the decay kinematics  can be solved event by event 
to obtain the sparticle masses in the decay cascade. 

Another important quantity is the $m_{T2}$ variable,
which is calculated from 
two visible momenta $p^{(i)}_{\rm vis} (i=1,2)$ 
and the missing transverse momentum ${\mathbf E}\slush_T$
as in Eq.(\ref{eq1}) \cite{Lester:1999tx,Barr:2003rg},
\begin{equation}
m_{T2}=
\min_{{\mathbf E}\slush_T={\bf p}\slush_{T}^{(1)}+{\bf p}\slush_{T}^{(2)}}
\Bigl[ \max \left\{
m_T(p^{(1)}_{\rm vis}, {\bf p}\slush_{T}^{(1)};M_{\rm test} )  ,
m_T(p^{(2)}_{\rm vis}, {\bf p}\slush_{T}^{(2)};M_{\rm test} )  
\right\}
 \Bigr].
\label{eq1}
\end{equation}
Here, $p^{(i)}_{\rm vis} (i=1,2)$ is the sum of momenta of the particles in the visible system $i$
which is a set of visible decay products from a parent particle $i$.
$M_{\rm test}$ is an arbitrary mass parameter 
called the test LSP mass.

The quantity is bounded above by the mass of the heavier of the initially 
produced sparticles if we set $M_{\rm test}=m_{\rm LSP}$. 
This property of the $m_{T2}$ is useful for determining the sparticle masses.
For example, for the process $pp\rightarrow \tilde{q}_R \tilde{q}_R\rightarrow jj 
\chi^0_{1}\chi^0_{1}$ studied in \cite{Weiglein:2004hn, LHCsusy08},   
the events are populated near the $m_{T2}$ endpoint,
which is very clearly visible and coincides with $m_{\tilde{q}_R}$.

It was pointed out recently that
 the endpoint of the $m_{T2}$ distribution $m^{\rm end}_{T2}$ 
 as a function of $M_{\rm test}$
has a kink at the true LSP mass in the case that 
the invariant mass of 
the visible system $m_{\rm vis}$ (which consists of jets and leptons)
can range \cite{Cho:2007qv,Gripaios:2007is,Barr:2007hy,Cho:2007dh,
Nojiri:2008hy,Barr:2008ba,Tovey:2008ui}.
The kink arises 
because the derivative of $m^{\rm end }_{T2}$ with respect to a test LSP mass 
 differs depending on the $m_{\rm vis}$,
 while 
$m_{T2}^{\rm end}$ cannot be above the parent sparticle 
masses $\max\{m_1,m_2\}\equiv M$ at $M_{\rm test}=m_{LSP}$, 
so every trajectory that an event makes on the $m_{T2}-M_{\rm test}$ plane
passes or goes below the point $(m_{T2},M_{\rm test})=(M,m_{LSP})$.
The LSP mass and gluino mass  may  be reconstructed 
from $M_{\rm test}$ and the $m_{T2}^{\rm end }$ value 
at the kink position. 
The mass determination has been demonstrated 
 in the  four jets and $E\slush_T$ channel  
at a certain MSSM model point
in which $\tilde{g}$ decays dominantly via
$\tilde{g}\rightarrow \tilde{q}q \rightarrow {\ti{\chi}}^0_1 qq$ 
and $\tilde q$ is heavy \cite{Cho:2007dh}.
This shows that short hadronic decay chains 
can also contribute to sparticle mass determination.

To make  
use of the SUSY events fully in  the early stage of the LHC, 
it is useful to define the $m_{T2}$ variable  in an inclusive manner
without any specification of decay modes. 
This is because a squark and a gluino 
may decay into a mode with more than two  jets in the final state. 
For example, the decay modes $\tilde{g}\rightarrow t\bar{t}\tilde{\chi}^0_1$ and  
$tb\tilde{\chi}^{\pm}_1$ 
have large branching ratios in large mSUGRA parameter regions, 
because the scalar top and the scalar bottom tend to have 
masses  much lighter than the first generation squark masses. 

In the previous paper, we  therefore define an inclusive  $m_{T2}$ variable using 
a hemisphere method  \cite{Nojiri:2008hy}.
The inclusive $m_{T2}$ is defined in two steps.
 In the first step,  we divide jets in each event into two hemispheres
  \cite{hemi, Matsumoto:2006ws}. 
This is normally done  by associating the jets with
 two leading axes which are initially taken 
as the two leading jets in the event.  
The sum of the jet momenta
assigned in a hemisphere is called a hemisphere momentum $p_{\rm hemi}^{(i)} (i=1,2)$. 
In the next step, 
a stransverse mass $m_{T2}$ is calculated using Eq.(\ref{eq1}) with 
$p_{\rm hemi}^{(i)}$ taken as $p_{\rm vis}^{(i)}$. 
The inclusive $m_{T2}$ as defined
above carries the information on the parent sparticle masses  $\max\{m_1,m_2\}$, 
if the hemisphere algorithm groups the decay products from the particle 1 
and 2 into two different hemispheres correctly. 
It is shown that a parent squark mass 
$m_{\tilde{q}}$ can be  obtained from  $m_{T2}^{\rm end}$ 
in the case of $m_{\tilde{q}}>m_{\tilde{g}}$.    
Moreover, the $m_{T2}^{\rm end}$ 
as a function of a test LSP mass still has a kink 
at the true LSP mass.  
The inclusive $m_{T2}$ distribution 
is also useful for discriminating model parameters
and discussed extensively in  \cite{Hubisz:2008gg}.

In this paper, we propose a \lq\lq sub-system $m_{T2}$'', $m_{T2}^{\rm sub}$ .
 It is defined as an inclusive $m_{T2}$ variable,  
but the highest $p_{T}$ jet is removed before the hemisphere reconstruction.   
The definition is inspired by an observation 
that the squark decays via $\tilde{q}\rightarrow\tilde{g}$ or  
$\tilde{q}\rightarrow \tilde{\chi}^{\pm }, \tilde{\chi}^0$ produce
a high $p_T$ jet if $m_{\tilde{q}}$ is sufficiently larger than 
$m_{\tilde{g}}$, $m_{\tilde{\chi}^{\pm}}$ and $m_{\tilde{\chi}^0}$. 
In the case that the jet from the squark decay is identified, 
the remaining system is either gluino-gluino or gluino-neutralino/chargino 
for $\tilde{q}\tilde{g}$ co-production,  
so  $m_{T2}^{\rm sub}({\rm  end})=m_{\tilde{g}}$.  
By studying several model points
we provide convincing cases 
that {\it both } $m_{\tilde{q}}$ and $m_{\tilde{g}}$ are estimated
using $m_{T2}^{\rm end}$ and $m^{\rm sub,end}_{T2}$. 
We also calculate background  distributions coming from the 
productions  $t\bar{t} + n$ jets, 
$W + n$ jets and $Z + n$ jets using {\tt ALPGEN} \cite{Mangano:2002ea,Mangano:2001xp} 
with MLM matching. 
We find that the signal to noise ratio ($S/N$) is large especially for the events 
near  the $m_{T2}$ endpoint, 
which are most sensitive to  squark and gluino masses. 
 
  The importance of
 matrix element (ME) corrections to  SUSY processes have been emphasized recently 
 \cite{Plehn:2005cq,talk_alwall, Alwall:2008ve}.
We provide an estimate of the size of the matrix 
element correction to the signal $m_{T2}$ distributions
using {\tt MadGraph} \cite{Alwall:2007st}. 
We find that  the signal $m_{T2}$ distributions 
 are not significantly modified by the SUSY matrix element corrections  near the 
endpoint of the  $m_{T2}$ distributions. 

This paper is organized as follows. In Section \ref{sec;partonlevel}, we describe the 
 sub-system  $m_{T2}$. 
We show  parton level $m_{T2}$ and $m^{\rm sub}_{T2}$ distributions,  
and discuss reconstruction efficiencies of the  SUSY decay cascades  
using a hemisphere algorithm at our sample model point.
The jet level distributions using {\tt HERWIG}  \cite{Corcella:2000bw}
with simple detector simulator {\tt AcerDET} \cite{RichterWas:2002ch} 
are given in Section \ref{sec;MC}. 
We study  the SM background distributions  
generated by {\tt ALPGEN} in Section \ref{sec;BG}.  
Section \ref{sec;conclusions} is devoted to the conclusions.

In Appendix \ref{sec;app_mt2},
we describe the condition that $m^{\rm end}_{T2}$ coincides with 
the mass of the heavier of the initially produced squarticles.
The effects of the matrix element corrections to 
$m_{T2}$ distributions are  studied in Appendix \ref{sec;app_ME}.

\section{The sub-system  inclusive $m_{T2}$ ($m^{\rm sub}_{T2}$) - Parton level analysis}
\label{sec;partonlevel}

At the LHC, squarks and gluinos are 
copiously produced via $\sq\sq, \sq\gluino$ and $\gluino\gluino$ production processes. 
Each of them decays into visible objects and a LSP.
If the visible systems are correctly grouped, 
the inclusive $m_{T2}$ with the correct $p_{\rm vis}^{(i)}$ defined as Eq.(\ref{eq1}) 
can be calculated.
In that case, the important property is\footnote{
The endpoint for $\sq\gluino$ production events 
is given by $\max\{m_{\sq},m_{\gluino}\}$ 
unless the LSP mass is too close to $m_{\gluino}$, 
which is satisfied in the typical mass spectrum.
More details are discussed in Appendix \ref{sec;app_mt2}.
}
\begin{eqnarray}
m_{T2}^{\rm end}=\max\{m_{1},m_{2}\}.
\end{eqnarray}
Here, $m_{T2}^{\rm end}$ denotes the endpoint of the $m_{T2}$ distribution
and $m_{1}$ and $m_2$ denote the masses of the produced parent particles.
In this section, the test mass is taken as the true LSP mass
($M_{\rm test}=m_{\rm LSP}$).

In the case of $m_{\sq} \gg m_{\gluino}$,
squark-gluino and gluino-gluino production events are 
dominant SUSY production processes.
They give the different $m_{T2}$ endpoints: $m_{\sq}$ and $m_{\gluino}$.
For the squark-gluino production events,
a squark decays dominantly into a gluino (or another lighter sparticle) 
and a jet.
If we can identify the jet,
all other elements of the system make up a sub-system
that may be considered as 
a gluino-gluino (or gluino-the other sparticle) system.
We introduce the variable $m_{T2}^{\rm sub}$ (sub-system $m_{T2}$),
which is $m_{T2}$ calculated for the sub-system.
The missing transverse momentum 
is taken as the same as for the whole system
since the sum of the two LSP momenta is required 
for the calculation of $m_{T2}$.
The expected endpoint of $m_{T2}^{\rm sub}$ is $m_{\gluino}$.

Practically, we define the  sub-system as the system with the highest $p_T$ jet removed.
If the highest $p_T$ jet is from a decay chain of a squark,
the endpoint of $m_{T2}^{\rm sub}$ is expected as,
\begin{eqnarray}
m_{T2}^{\rm sub,end}=\min\{m_1,m_2\}.
\end{eqnarray}

We now show parton level 
 $m_{T2}$ and  $m^{\rm sub}_{T2}$  distributions at  
 our  sample model points. Here,   
 we take the model points \lq\lq a - f'' listed in Table \ref{model} with 
 the  GUT scale gaugino mass $M_{1/2}=300$~GeV and  $\tan\beta=10$. 
  The GUT scale sfermion mass  $m_0$  is   600~GeV at point f 
  and  1400~GeV at point a. The gluino masses at   
 these points are  approximately the same, while squark masses range 
 from $\sim 900$~GeV (at point f)  to $\sim 1500$~GeV (at point a).
The  GUT scale Higgs masses are tuned  so that 
 the $\mu$ parameters are small $\sim 180$~GeV. The relation
  $\mu \sim M_1$ results in a thermal relic density of the LSP 
that is consistent with the observed 
 cold dark matter density of our universe \cite{Drees:2000he, Ellis:2002wv, 
 Ellis:2002iu}\footnote{The choice of 
 the $\mu$ parameter does not affect $m_{T2}$ distributions 
 discussed in this paper.}. 
Some of the branching ratios of 
the 1st generation squarks are given in Table \ref{modelbr}.
 We calculate the masses and the branching ratios at these  model points 
 using {\tt ISAJET} \cite{Paige:2003mg}, and the mass parameters are 
 interfaced to {\tt HERWIG} \cite{Corcella:2000bw} 
 using {\tt ISAWIG} \cite{isawig}. The cross sections are calculated 
 using {\tt HERWIG}. 
  Note that the values of $m_0$ and $M_{1/2}$ in Table \ref{model} 
are within the discovery reach in mSUGRA for the  ATLAS and CMS experiments 
at $\int dt L =1\,$fb$^{-1}$.

\TABLE[!ht]{
\begin{tabular}{|c||c|c|c|c|c|c|}
\hline
& $m_0$& $A_0$ & $m_{\tilde{q}}$& $m_{\tilde{g}}$&  $m_{\rm LSP}$  & $\mu$ \\
\hline
a & 1400& $-1400$&   1516 & 795.7 & 107.9  &  180\\
b & 1200 & $-1200$ &  1342 & 785.0 & 107.4 &  180\\
c & 1100 & $-1100$ & 1257 & 779.5 & 107.1  &  180\\
d & 1000 & $-1000$ &  1175 & 773.2 & 106.8 &  180\\
e &  820 &$-750$ &    1035 & 761.7 & 106.1 &  180\\
f &  600 &$-650$ &  881.0 & 745.4 & 107.8  &  190\\
\hline
\end{tabular}
\caption{Some of the mass parameters of  our model points. 
We take the scalar masses of sfermions and gaugino masses to be universal. 
We tune the higgsino mass parameter $\mu$ by allowing 
non-universal 
GUT scale Higgs masses parameters 
so that  $\Omega h^2 \sim 0.1$.  All mass parameters are given in GeV.  
}\label{model}
}

\TABLE[!ht]{
\begin{tabular}{ |c |c |c |c|c|c| }
\hline
point &$m_{\tilde{u}_L}$ & $ Br(\tilde{u}_L\rightarrow \tilde{g} u)$ &$Br(\tilde{u}_R\rightarrow \tilde{g}u) $ &  $\sigma$(SUSY)(pb)  & $\sigma(\tilde{q})$ (pb) \\
\hline 
a  &    1516   &      0.71   &           0.93 & 4.91& 0.46\\
b  &   1342   &        0.68    &        0.92  &5.35 &  0.79\\
c  &  1257    &        0.66   &        0.91 &5.84  & 1.07\\
d &1175  &     0.62 &       0.90  & 6.15 & 1.40\\ 
e &1035  &     0.53  &       0.96 & 7.31& 2.36\\
f  &  881 &         0.31 &       0.71 &9.49 &  4.34\\
\hline 
\end{tabular}
\caption{Some  relevant branching ratios of squarks  are calculated using {\tt ISAJET}.  
The total SUSY production cross section and the cross section involving at 
least one first generation squark  estimated using  {\tt HERWIG} are also given. }
\label{modelbr}
}

The squark production cross section reduces quickly 
with increasing first generation squark masses. 
The total SUSY production cross section varies
more than a factor of two from point a  to point f.
The difference comes mostly from the decrease of $\sigma(\tilde{q}\tilde{g})$. 
In particular, the production cross section involving at least one first generation 
squark  is only  0.46~pb at point a  and 4.34~pb at point f.  
The gluino-gluino production cross section also becomes reduced 
because $t$-channel squark exchange is suppressed.
Chargino and neutralino production is important at point a.

The squark decays dominantly into gluino  and a jet (See Table 2). 
The squark branching ratio 
into the gluino is dominant except at point f.
For points a to c,
the mass difference between squark and 
gluino is significantly larger than half of the gluino mass.
Therefore, a jet from a  squark decay is likely  to be the highest $p_T$ jet
in the  event. 
Jets from the other squark decay modes such as 
$\tilde{q}\rightarrow \tilde{\chi}^0_i j$ 
and  $\tilde{\chi}^\pm j$ 
have  $p_T$  which is even higher than that of $\tilde{q}\rightarrow\tilde{g}j $ 
on average.

To define the inclusive $m_{T2}$ and $m_{T2}^{\rm sub}$ distributions, we group 
jets  in an event into two \lq\lq visible objects''.   
For this purpose, we adopt the hemisphere method 
in Refs \cite{hemi, Matsumoto:2006ws}.
For each event, two hemispheres are defined and high $p_T$
 jets are assigned to one of the hemispheres 
 as follows:
\begin{enumerate}
\item Each hemisphere is defined by an axis $p_{\rm hemi}^{(i)}$ ($i = 1, 2$), 
which is the sum of the momenta of the selected  high $p_T$ objects 
belonging to the hemisphere $i$. \item  A high $p_T$ object $k$ belonging to a hemisphere $i$ 
satisfies the following conditions: 
\begin{equation}
d(p_k, p_{\rm hemi}^{(i)}) <d(p_k, p_{\rm hemi}^{(j)}), 
\end{equation}
where the function $d$ is defined by 
\begin{equation}
d(p_k, p_{\rm hemi}^{(i)})= (E_{\rm hemi}^{(i)}-\vert {\bf p}_{\rm hemi}^{(i)} \vert  \cos\theta_{ik})
 \frac{E_{\rm hemi}^{(i)}}{(E_{\rm hemi}^{(i)}+E_k)^2}
\end{equation}
Here $\theta_{ik}$ is the angle between ${\bf p}_{\rm hemi}^{(i)}$ and ${\bf p}_k$. 
\end{enumerate}

For our  $m_{T2}$ analysis in this paper,  
the selected objects are  jets with 
$p_{Ti} > 50$~GeV  and $\vert \eta_i  \vert <3$. We do not include the 
jets with $p_{T}\le 50$~GeV nor $\vert\eta_i\vert>3$ to avoid contaminations 
from initial 
state radiations.  For $m^{\rm sub}_{T2}$, we do not include the 
highest $p_{T}$ jet in the selected objects. 

To find the $p_{\rm hemi}^{(i)}$, we adopted the algorithm discussed in Refs. 
\cite{hemi}: 
\begin{enumerate}
\item We first take the highest $p_T$ object with momentum 
$p_1$ 
and the object  $i$ which 
has the largest $ p_{Ti} \times \Delta R(1,i)$, where 
$\Delta R(i,j)  =\sqrt{(\eta_i-\eta_j)^2 +(\phi_i-\phi_j)^2}$. 
\item 
We regard $p_{1}$ and $p_i$ as two seeds of the 
initial hemisphere axes, and assign the other objects to one of the 
two axes. 
\item We recalculate the hemisphere axes.  
We  perform iterations until the 
assignment converges.  
Once $p_{\rm hemi}$'s are determined, 
one can calculate $m_{T2}$ by using Eq.\ref{eq1} 
with taking $p_{\rm hemi}$ as $p_{\rm vis}$. 
\end{enumerate}

In this section, we study {\it parton level events}. 
The momenta of quarks and gluons from sparticle
decays are extracted from {\tt HERWIG}
 event records,  and only $\tilde{q}$-$\tilde{q}^{(*)}$, $\tilde{q}^{(*)}$-$\tilde{g}$
 and $\tilde{g}$-$\tilde{g}$ 
productions are included in the figures. 
When a sparticle decays into 
$W^{\pm}$,$ Z^0$, and $t$, we further follow their decays. 
Note that each parton is in general off-shell when 
they are created from a sparticle decay,  
and we do not follow  parton shower evolutions after that.
We  do not include partons from initial state radiations. 
These effects will be taken into account in particle level MC  simulations 
in the next section. 

For the calculation of the $m^{\rm sub}_{T2}$,
we remove the highest $p_{T}$ jet before the hemisphere assignment
to obtain the $p_{\rm hemi}$'s.
As an alternative definition, 
we can remove the highest $p_T$ jet from the $p_{\rm hemi}$'s 
after the hemisphere assignment,
and $\tilde{m}^{\rm sub}_{T2}$ denotes this alternative $m_{T2}$ in the following.

We now compare $m_{T2}^{\rm sub}$ and $\tilde{m}_{T2}^{\rm sub}$ 
at point b in Fig.~\ref{Figure:mt2subp3}.  
In the left plot, we show the $m^{\rm sub}_{T2}$ distribution in 
the solid line. 
In the right plot, the solid line shows the $\tilde{m}_{T2}$ distribution.
In each plot, the dotted line shows the \lq true distribution'
$m_{T2}^{\rm sub}({\rm true})$, 
in which the $p_{\rm vis}^{(i)}$ consists of 
the momenta of decay products from a parent particle $i$ except for the highest $p_T$
parton using the generator information.
This is an ideal distribution 
when the assignment of the visible systems is perfect.
Note that the highest $p_T$ jet is not always from a $\sq$ decay.
Even in the distribution of $m_{T2}^{\rm sub}({\rm true})$,
two endpoints can be seen, 
the lower is at the gluino mass and the 
higher is at the squark mass. 

\FIGURE[!ht]{
\includegraphics[width=5cm]{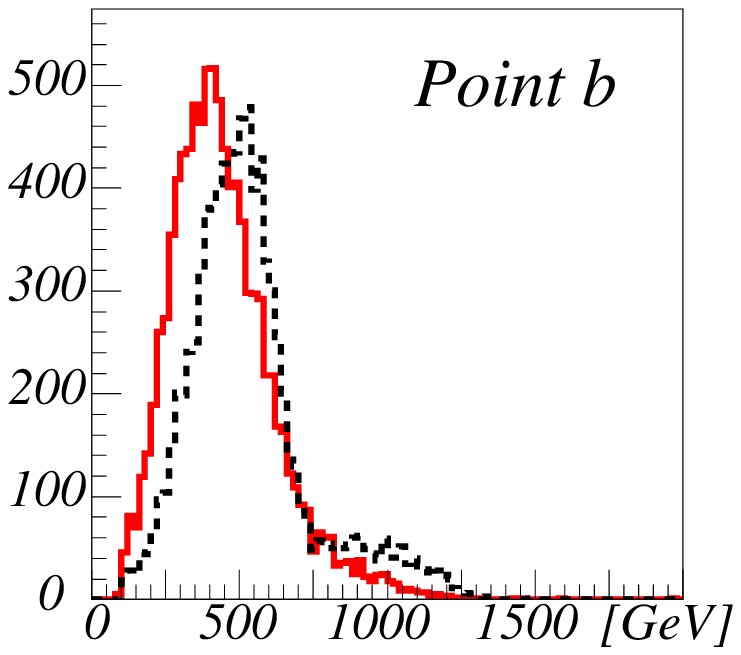}
\includegraphics[width=5cm]{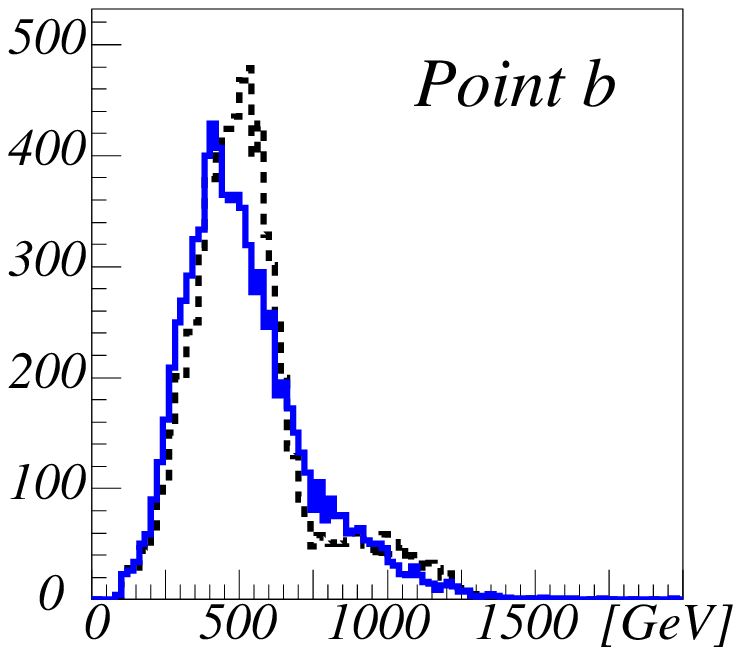}
\caption{Various parton level $m^{\rm sub}_{T2}$ distributions at point b. 
The left plot,  a solid line: the $m_{T2}^{\rm sub }$ distribution.
The right plot, a solid line: a $\tilde{m}^{sub}_{T2}$ distribution, 
which is a $m_{T2}$ variables but calculated after 
subtracting the highest $p_T$ object from the hemisphere momenta. 
Dotted lines,  $m^{\rm sub}_{T2}({\rm true})$ distributions, which 
uses  generator information for the hemisphere assignment. 
Arbitrary normalizations are used for the $y$-axes.
See text for details.
} 
\label{Figure:mt2subp3}
}

The endpoint at the gluino mass 
is more clearly visible for the $m^{\rm sub}_{T2}$ 
than for the $\tilde{m}^{\rm sub}_{T2}$ distribution.  
The improvement in $m^{\rm sub}_{T2} $ distribution  
may be explained as follows.  
At point b, a parton from 
$\tilde{q}\rightarrow q \tilde{g}$ has
a large open angle to the gluino decay products on average.
The event effectively has three axes:
the two momenta of the two gluino decay products 
and the momentum of the extra parton from squark decay.
The assumption of the hemisphere algorithm 
that events must have two axes 
may lead to an incorrect hemisphere assignment.
Removing the highest $p_T$ parton 
before the hemisphere assignment 
therefore makes the kinetic endpoints more visible.  

The successful endpoint reconstruction 
shows  that the hemisphere algorithm reconstructs  
a  total visible momentum  of a  squark/gluino decay 
more or less correctly. 
One can check this explicitly 
by counting the 
number of partons assigned to an incorrect hemisphere. 
(Fig.~\ref{assign}).  
The solid (dashed) histograms correspond to the 
distributions of the number of mis-reconstructed partons 
for the case 
that the highest $p_T$ parton is removed
before (after) the hemisphere assignment.
The improvement achieved by removing the highest $p_T$ parton before the hemisphere assignment
is clearly seen at point a (the left plot).  
At this point, 
$m_{\tilde{q}}=1516$~GeV and $m_{\tilde{b}}=796$~GeV,
so the parton from the squark decay should have $p_T$ of the order of several hundred GeV.
We also see mild improvement at point b (the middle plot).
At point e (the right plot), 
the squark and gluino masses are close,
$(m_{\tilde{q}}-m_{\tilde{g}})/ m_{\tilde g} =0.36$. 
In this case, removing the highest $p_T$ jet before the hemisphere assignment
leads to the slightly worse reconstruction efficiency. 
The number 
of mis-reconstructed partons is either 0 or one for 
 more than half of the events in Fig.~\ref{assign}. 
  
\FIGURE[!ht]{
\includegraphics[width=4cm]{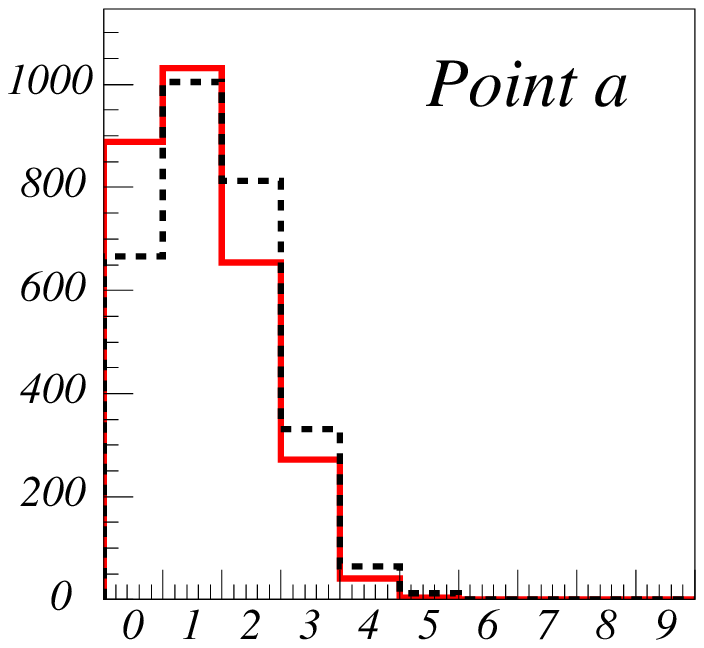}
\includegraphics[width=4cm]{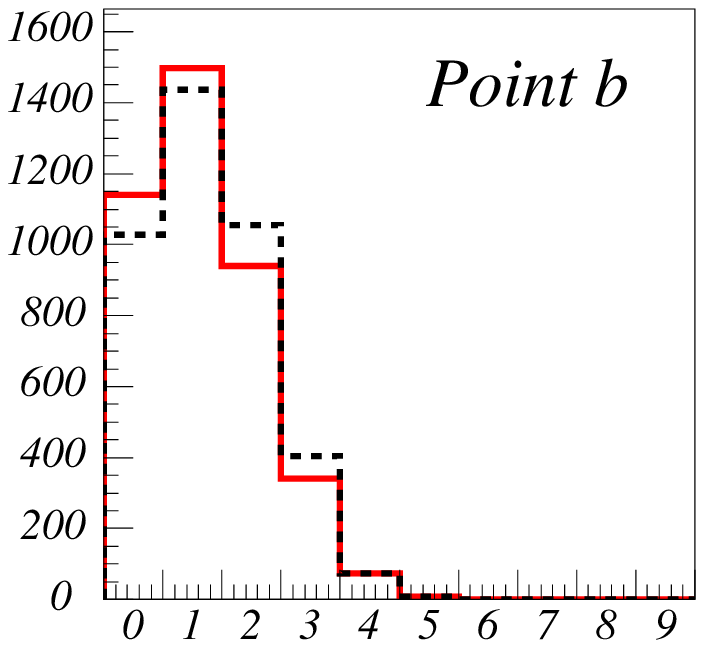}
\includegraphics[width=4cm]{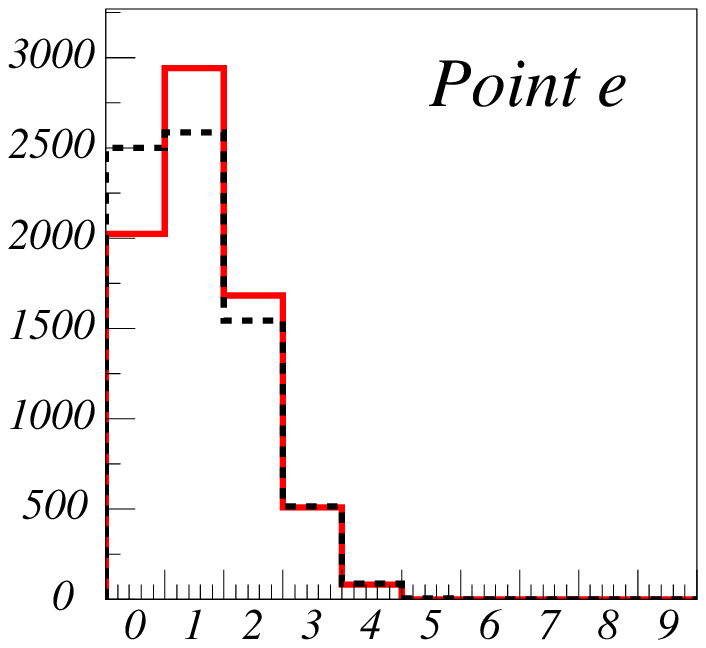}
\caption{The distributions of the number of partons assigned 
to an incorrect hemisphere
at point a (left), b (middle)
and e (right). 
In all figures, solid (dashed) histograms correspond to  
the case that the highest $p_T$ parton is removed before (after) 
the hemisphere assignment. 
Arbitrary normalizations are used for the $y$-axes.
See text for details.} 
\label{assign}
}

\FIGURE[!ht]{
\includegraphics[width=5cm]{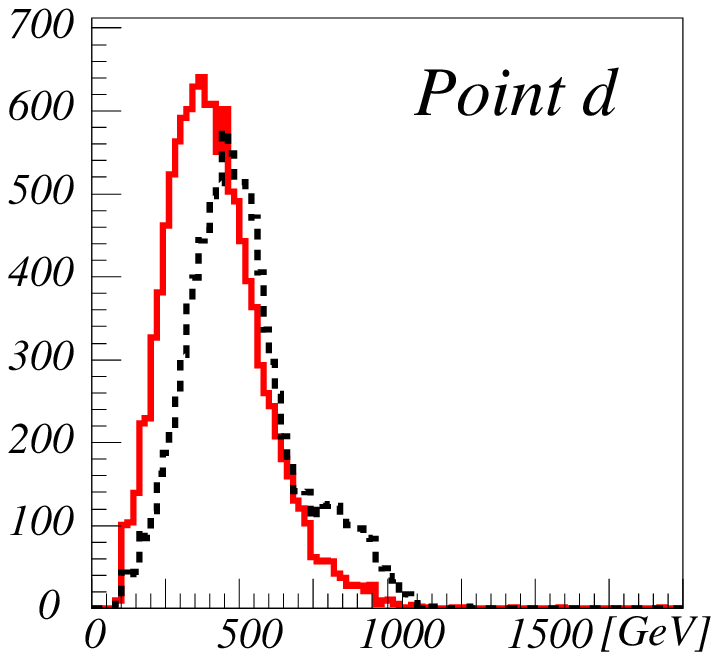}
\includegraphics[width=5cm]{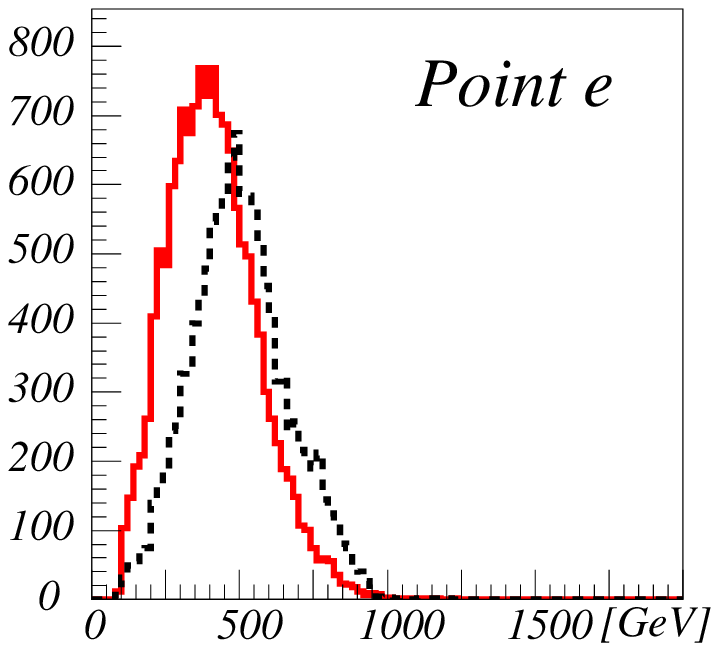}
\caption{The  $m^{\rm sub}_{T2}$ distributions (solid) 
and the $m^{\rm sub}_{T2}({\rm true})$ distributions (dashed)   
for squark-gluino co-production events at points d and e.  
Arbitrary normalizations are used for the $y$-axes.
 }\label{submt2de}
}

If the highest $p_T$ parton does not arise from $\tilde{q}$ decay, 
the gluino endpoint cannot be reconstructed 
even for squark-gluino production events. 
The probability  
strongly depends on the model parameters. 
In Fig.~\ref{submt2de} we show the $m^{\rm sub}_{T2}$ distributions 
for only squark-gluino co-production 
events at points d (left) and e (right). 
The gluino endpoint can be seen around  750~GeV 
from the $m_{T2}^{\rm sub}$ distribution at point d, 
which is close to that of the $m_{T2}^{\rm sub}({\rm true})$ distribution shown in the dotted line. 
However, at point e, even in the $m_{T2}^{\rm sub}({\rm true})$ distribution 
we cannot see the clear structure at the gluino mass. 

The difference between  the  $m_{T2}^{\rm sub}$  and 
$m_{T2}^{\rm sub}({\rm true})$ distributions at  points d and e  may be explained  as follows. 
At point d, $(m_{\tilde{q}}-m_{\tilde {g}})/m_{\tilde{g}}= 0.52$, 
and the energy of the parton from the squark decay is  bigger than that from the 
gluino on average. 
This is why $m^{\rm sub}_{T2}$ shows 
clear gluino endpoints at point d. 
In contrast, $(m_{\tilde{q}}-m_{\tilde {g}})/m_{\tilde{g}}= 0.36$   
at point e.  
The $m_{\sq} - m_{\gluino}$ is not large enough, 
and it is not likely that the parton from $\tilde{q}
\rightarrow \tilde{g}q$ has significantly high $p_T$ compared with 
those coming  from the gluino decays.  
This is why the $m^{\rm sub}_{T2}({\rm true})$ distribution  
does not show the endpoint at the gluino mass, 
it could be a problem to extract the gluino mass 
from the $m^{\rm sub}_{T2}$ distribution. 
However, 
the $m_{T2}^{\rm sub} $ distribution of $\tilde{q}\tilde{g}$  production 
is significantly smeared towards the lower $m_{T2}$ value. 
In the actual situation, 
the contribution from gluino-gluino pair productions 
would be added, and the $m_{T2}^{\rm sub}$ distribution 
would have the endpoint at the gluino mass. 
We will see in the next section that 
the contamination from squark-gluino  production is not serious. 

\FIGURE[!ht]{
\includegraphics[width=5cm]{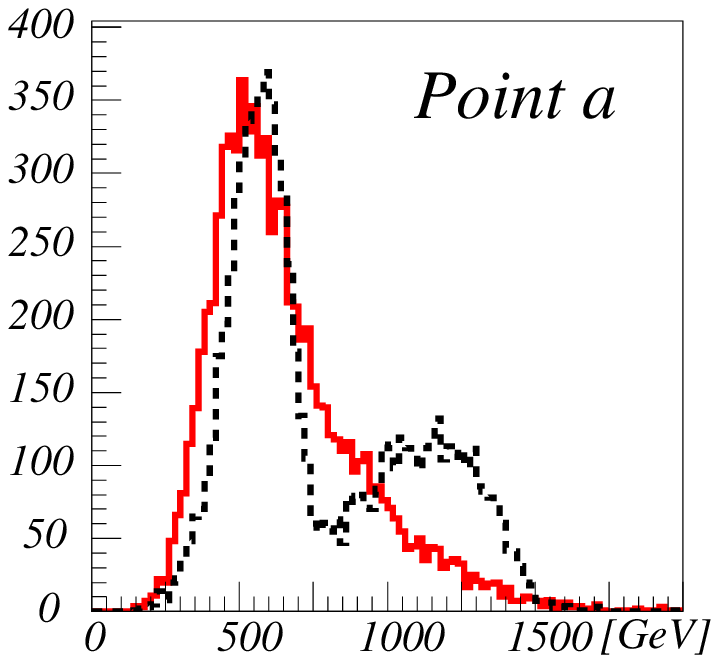}
\includegraphics[width=5cm]{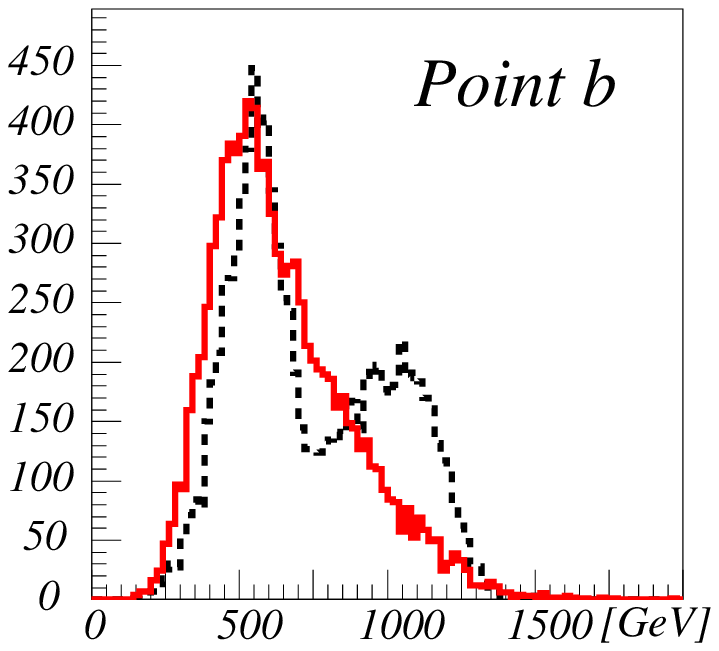}
\includegraphics[width=5cm]{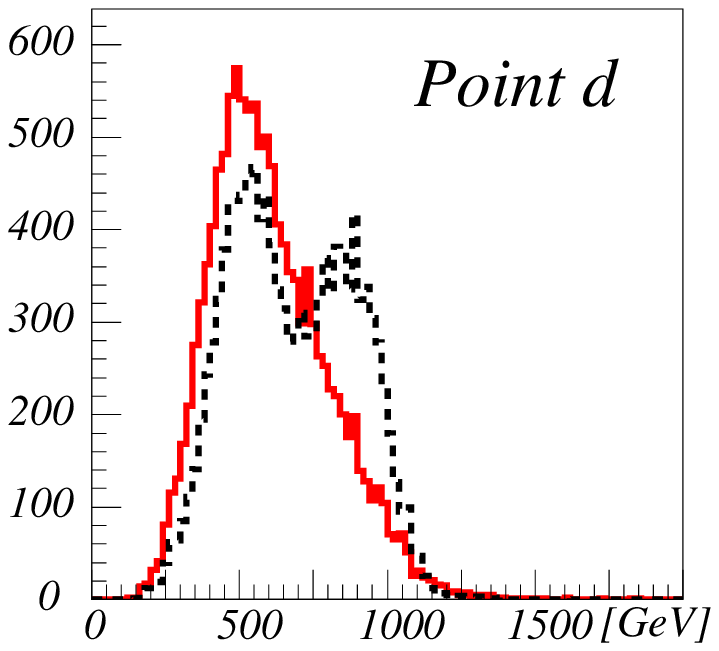}
\includegraphics[width=5cm]{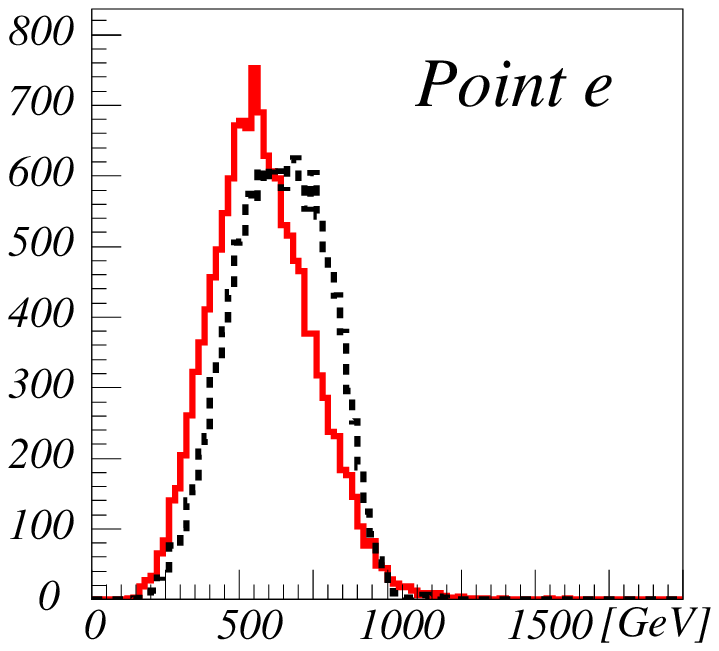}
\caption{The parton level  $m_{T2}$ distributions at points a, b, d and e.  
The solid  lines correspond to  the distributions using the hemisphere algorithm, while 
the dotted lines correspond to those  with  correct  parton assignments
obtained by using the generator information.
Arbitrary normalizations are used for the $y$-axes.
}\label{Figure:rmt2}}

For completeness, 
we show the parton level $m_{T2}$ distributions at points a, b, d and e
in Fig.~\ref{Figure:rmt2} to emphasize the difference between $m_{T2}$ and 
  $m^{\rm sub}_{T2}$ distributions. 
The solid histograms are the $m_{T2}$ distributions using the hemisphere algorithm,
 while the dotted  histograms correspond to the $m_{T2}({\rm true})$ 
distributions which are obtained by assigning the partons arising 
from a parent particle $i$ to hemisphere $i$ using generator information.  
At points a, b and d, 
the $m_{T2}({\rm true}) $  distribution has two peaks. 
The peak at the lower $m_{T2}$  value comes from gluino-gluino production,
while the  peak at higher $m_{T2}$ corresponds to the 
squark-gluino and squark-squark productions.  The endpoint of the 
distributions coincides with squark mass. 
The double peak structure cannot be 
seen in the distributions of $m_{T2}$, but the  endpoints are the
 same as that of $m_{T2} ({\rm true})$. 

The slope of the 
distribution near the endpoint becomes flatter with increasing squark mass
as can be seen from the distributions at points a, b, and d. 
In particular, the existence of a high $p_T$ parton 
from squark decay leads to some confusions in the hemisphere algorithm at point a, and 
a careful study of the distribution would be required to extract the 
squark mass from the fit.
The peak of the $m_{T2}$ distribution  coincides
with the lower $m_{T2}({\rm true})$ peak.  
The events near the peak 
come from   gluino pair productions  at  points a, b, and  d.  
In principle, the position of the peak contains the gluino mass information.
However,
this is not easy to observe because  the SM background 
may also be large in this region. 
At point  e,  although the endpoints of the $m_{T2}$ and $m_{T2}({\rm true})$ distributions are consistent,
the squark and gluino masses are 
too close for the two peak structure to be seen.

\section{The MC simulation of the signal}
\label{sec;MC}
We have shown that the endpoints of $m_{T2}$ and $m_{T2}^{\rm sub}$ 
distributions carry the information on squark and gluino masses using 
 parton level events. 
In this section we study the events produced 
by a parton shower Monte Carlo {\tt HERWIG} 
(in the particle level)  with a detector simulator 
{\tt AcerDET} under the 
 set of cuts to reduce the standard model backgrounds. 
The simple snowmass cone algorithm implemented in {\tt AcerDET} is used for
finding jets and we set the cone size $R=0.4$.
\FIGURE[!ht]{
\includegraphics[width=4.4cm]{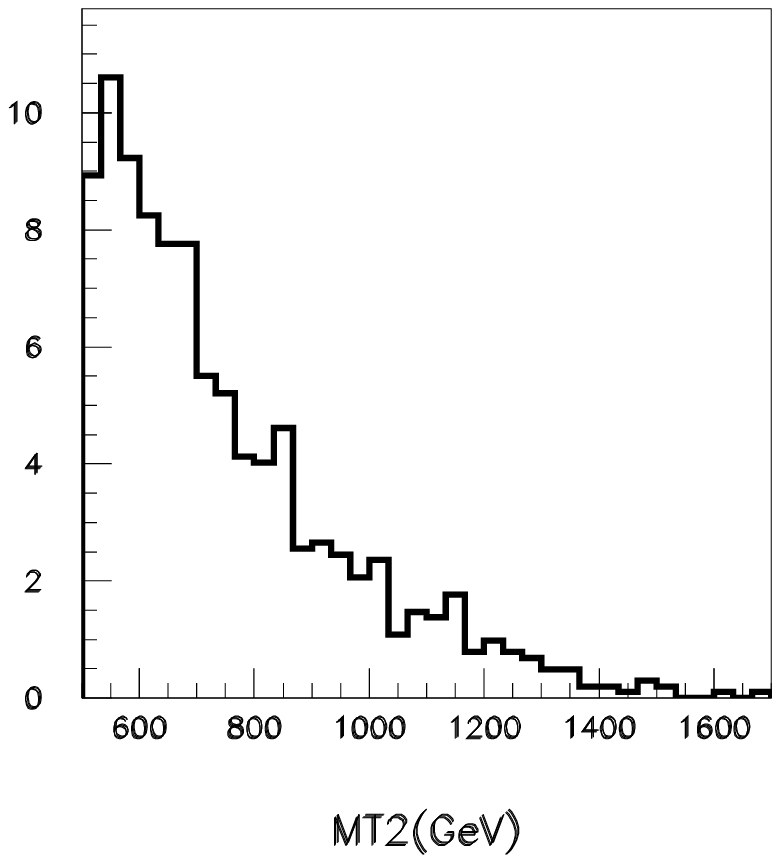}\hskip -1cm
\includegraphics[width=4.4cm]{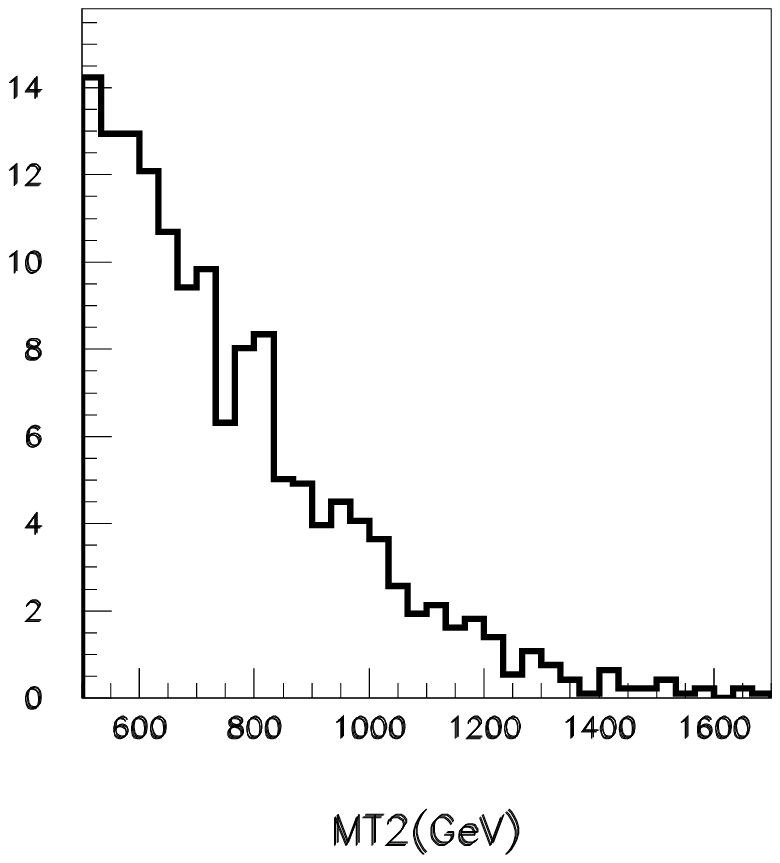}\hskip -1cm
\includegraphics[width=4.4cm]{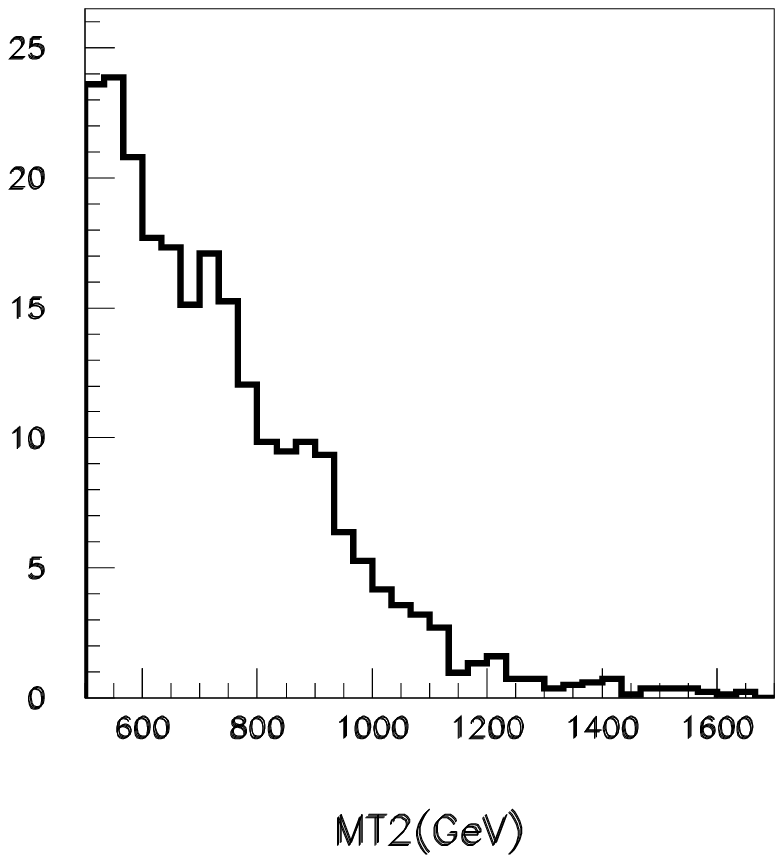}\hskip -1cm
\includegraphics[width=4.4cm]{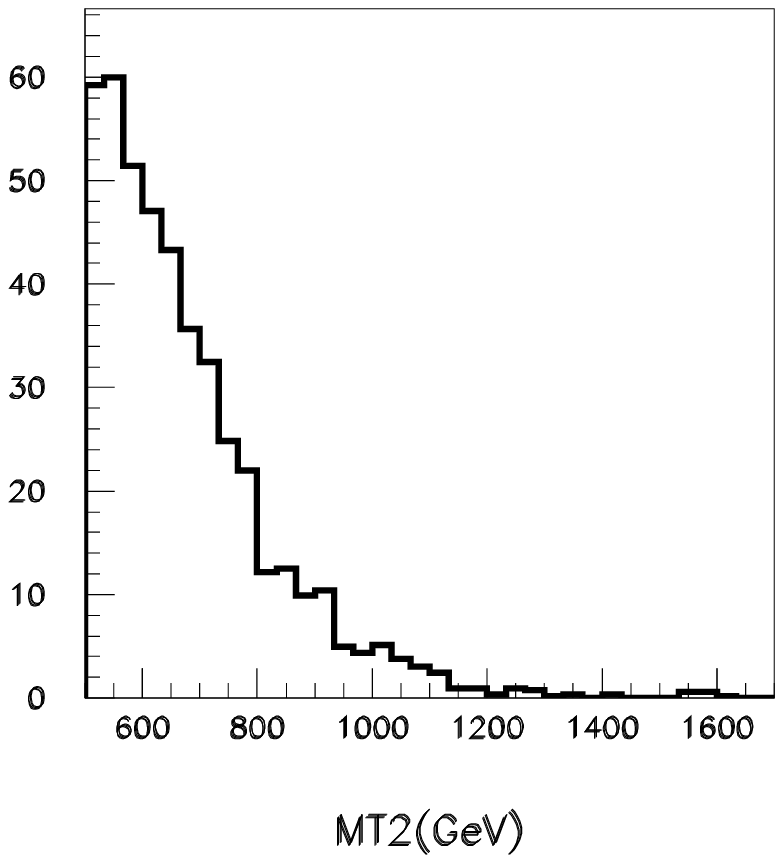}
\caption{The inclusive $m_{T2}$ destributions  at points a b, 
d and  f (from left to right).  
The up squark mass is  1516, 1342,  1175, 881~GeV from 
left to right, respectively.
Unit of $y$-axes is events/bin/$1\,$fb$^{-1}$.
}
\label{Figure:rmt2sb_1fb}}

\FIGURE[!ht]{
\includegraphics[width=7cm,angle=90]{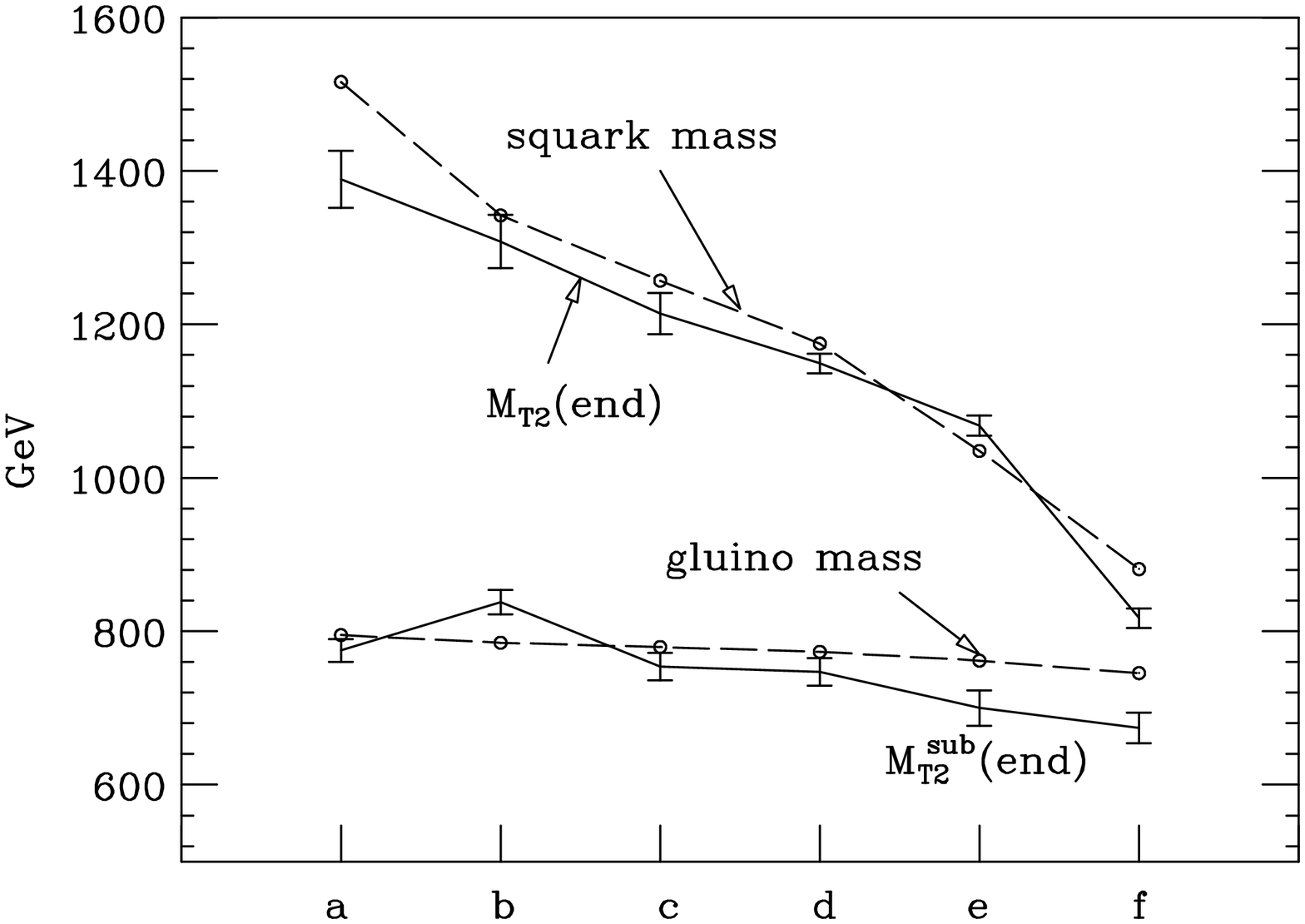}
\caption{The fitted  $m_{T2}$ and $m_{T2}^{sub}$ endpoints (solid lines)  
and  $m_{\tilde{q}}$ and $m_{\tilde{g}}$ (dashed lines) at each model point.
The bars show the size
of statistical errors for 50,000 SUSY events.
}
\label{Figure:mt2fit}
}

We apply the following cuts to the events. 
\begin{itemize}
\item Jet $p_T$ cuts: $n_{50}\equiv N_{\rm jet}(p_T>50~{\rm GeV} ) \ge 4$, 
$n_{100}\equiv N_{\rm jet}(p_T>100~{\rm GeV}) \ge 1$. 
\item $M_{\rm eff}(\equiv \sum_{p^{\rm jet}_T>50 {\rm GeV} }p_T + E\slush_{T})
>500$~GeV  
\item Transverse sphericity: $S_T>$ 0.2.
\item Missing Transverse momentum:$E\slush_{T}>200$~GeV, $E\slush_{T}> 0.2 M_{\rm eff}$. 
\item No isolated lepton with $p_T>20$~GeV. 
\end{itemize}
These cuts are similar to the standard SUSY cuts in the ATLAS TDR \cite{Atlas}, 
except for our tighter $E\slush_{T}$  cut.  
We veto events with isolated leptons because a hard lepton might be 
assoicated with a hard  neutrino. If there is a hard neutrino in an event, 
 $p_{\rm Tmiss}$  of the event may  not be the sum of the  transverse 
momenta of LSPs. In that case, the endpoint of the $m_{T2}$ distribution 
might be smeared.  

We first show the $m_{T2}$ distributions for our model points.
The $m_{T2}$ distributions for $m_{\rm test}= 10$~GeV under 
the SUSY cuts are shown in Fig.~\ref{Figure:rmt2sb_1fb}\footnote{ 
We set $m_{\rm test}$ small 
as we do not know the LSP mass initially.}. 
For each point, we have generated 50,000 SUSY events and the distribution 
is scaled to correspond to 
$\int dt {\cal L}=1\,$fb$^{-1}$ of luminosity. 

The endpoint of the $m_{T2}$ distribution is roughly at $\sim m_{\tilde{q}}$. 
We fit the distributions  to  linear functions 
\begin{eqnarray}
f(m)&=&a (m-m^{\rm end}_{T2}) +c  \ \ ({\rm for } \ \ m<m^{\rm end}_{T2})
 \cr
 &= & b(m-m^{\rm end}_{T2} ) +c   \ \  ({\rm for} \ \  m> m^{\rm end}_{T2}),
\end{eqnarray}

and the fitted $m^{\rm end}_{T2}$ values
are shown in Fig.~\ref{Figure:mt2fit}. 
Here, the statistical errors shown in bars 
correspond to 50,000 total SUSY events. 
The obtained $m_{T2}^{\rm end}$ and  $m_{\tilde{q}}$ are consistent 
except at points a and f.  
For point f, the squark and gluino masses 
are too close, and it is natural that the endpoint fall at weighted 
mean of gluino and squark masses.  
For point a, 
due to the very large mass difference between squark and gluino,  the 
hemisphere method involving the highest $p_T$ jet does not work perfectly.

Note that there is some ambiguity in choosing a fitting region. 
For example, for point a,  
the distribution consists of the two components, 
one arising from the gluino-gluino production 
with the endpoint around $800\,$GeV and the other from the squark-gluino production 
with the endpoint around $1400\,$GeV. We fit the distribution above 
$m_{T2}> 1000\,$GeV for this point. 
If we did the same fit at point f (the right plot), 
we might fit the mis-reconstructed tail of the events 
and therefore might obtain the endpoint at $1150\,$GeV.
This suggests that the region of the fit must be chosen carefully.  
In particular, the events near the fitted endpoint must make up a sizable fraction of 
the total events.
For points b, d and f, 
we first fit the region from the $m_{T2}$ slightly above the peak position of the distribution 
 up to the highest bin with enough statistics ($>10$ events/bin).
We then increase the lower limit until we obtain a small $\Delta\chi^2$. 
The $\Delta \chi^2/{\rm n.d.f}$  is less than 1 except at points c and e, and  
all fits satisfy $\Delta \chi^2/{\rm n.d.f}<2$. 

\FIGURE[!ht]{
\includegraphics[width=6cm]{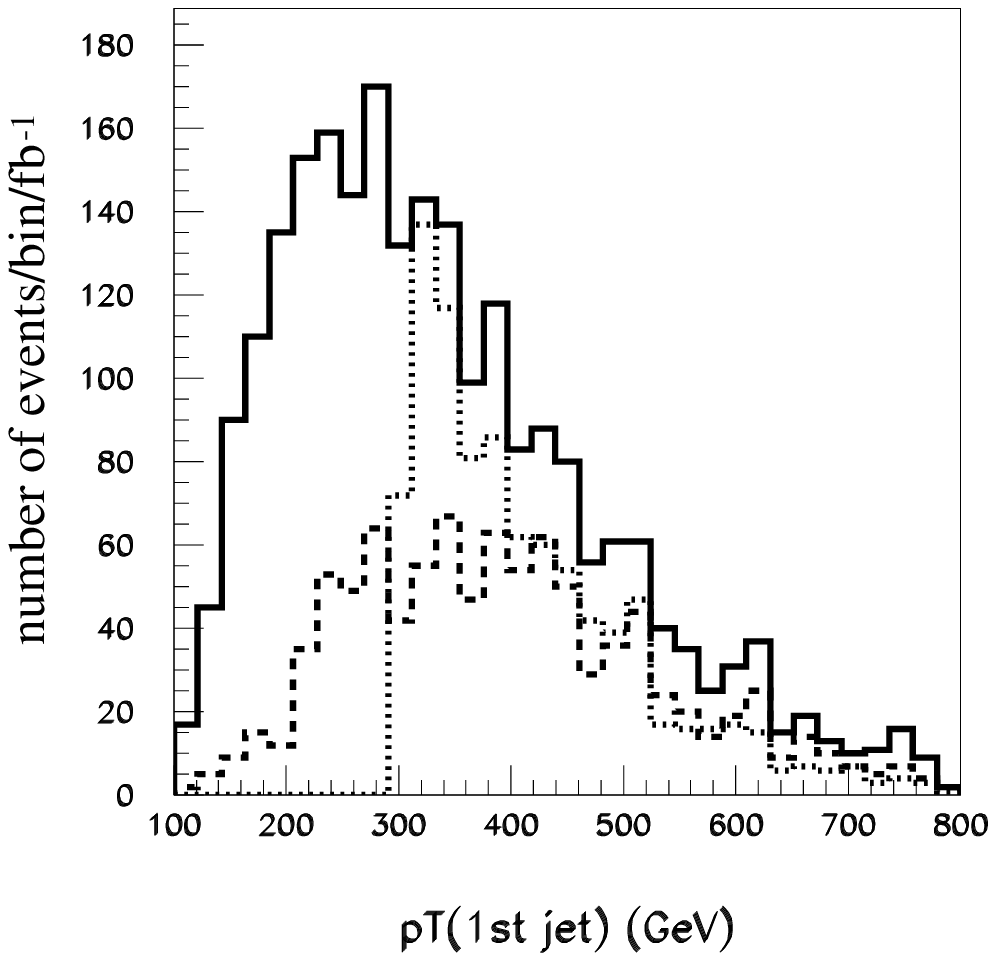} \hskip -1cm 
\includegraphics[width=6cm]{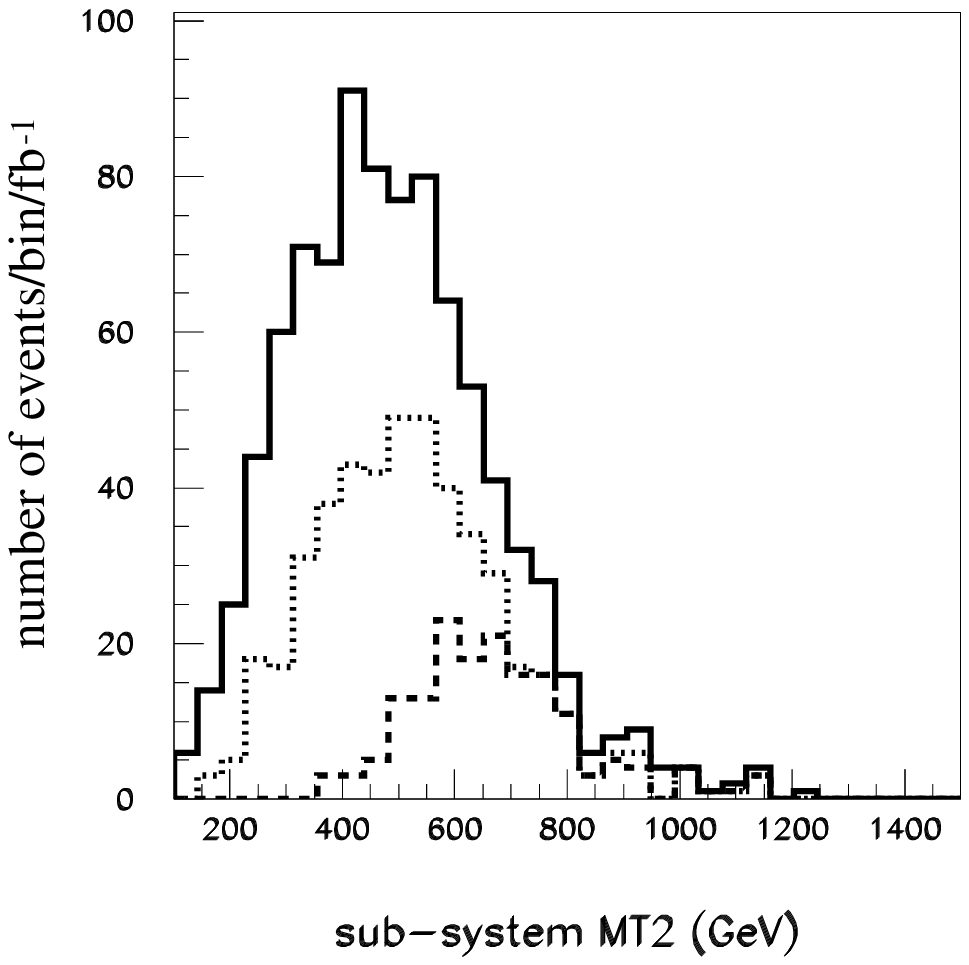} \hskip -1cm 
\caption{The $p_T$ distributions of the highest $p_T$ jet (left) among 
the jets with $\vert \eta \vert <3$, and the distributions of the 
 $m^{\rm sub}_{T2}$ (right) at point a. 
The dashed lines show the contributions from the events with $n_{300}=1$
and the dotted lines show the contributions from the events with $n_{\sq}=1$.
The standard SUSY cuts are applied for all plots, and $m_{\rm test} = 10$~GeV.
Unit of $y$-axes is events/bin/$1\,$fb$^{-1}$.
}
\label{Figure:pt1jets}
}

\FIGURE[!ht]{
\includegraphics[width=6cm]{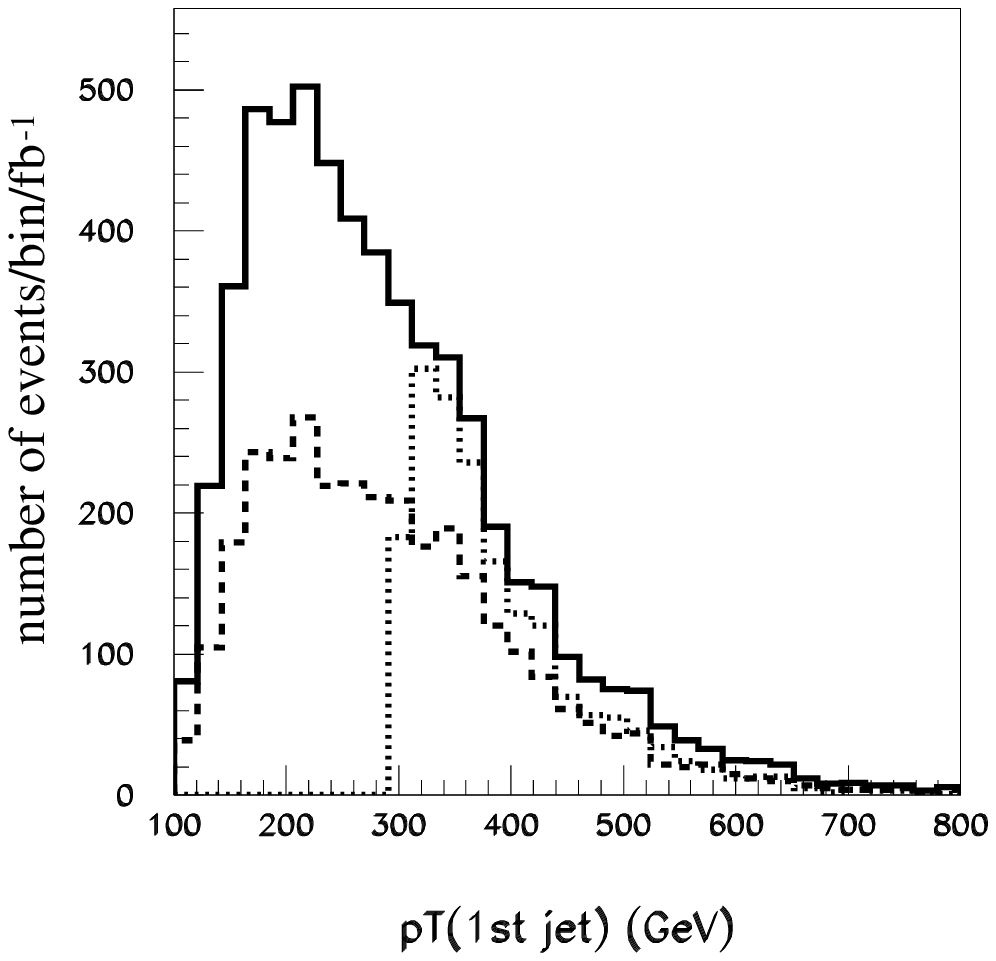} \hskip -1cm
\includegraphics[width=6cm]{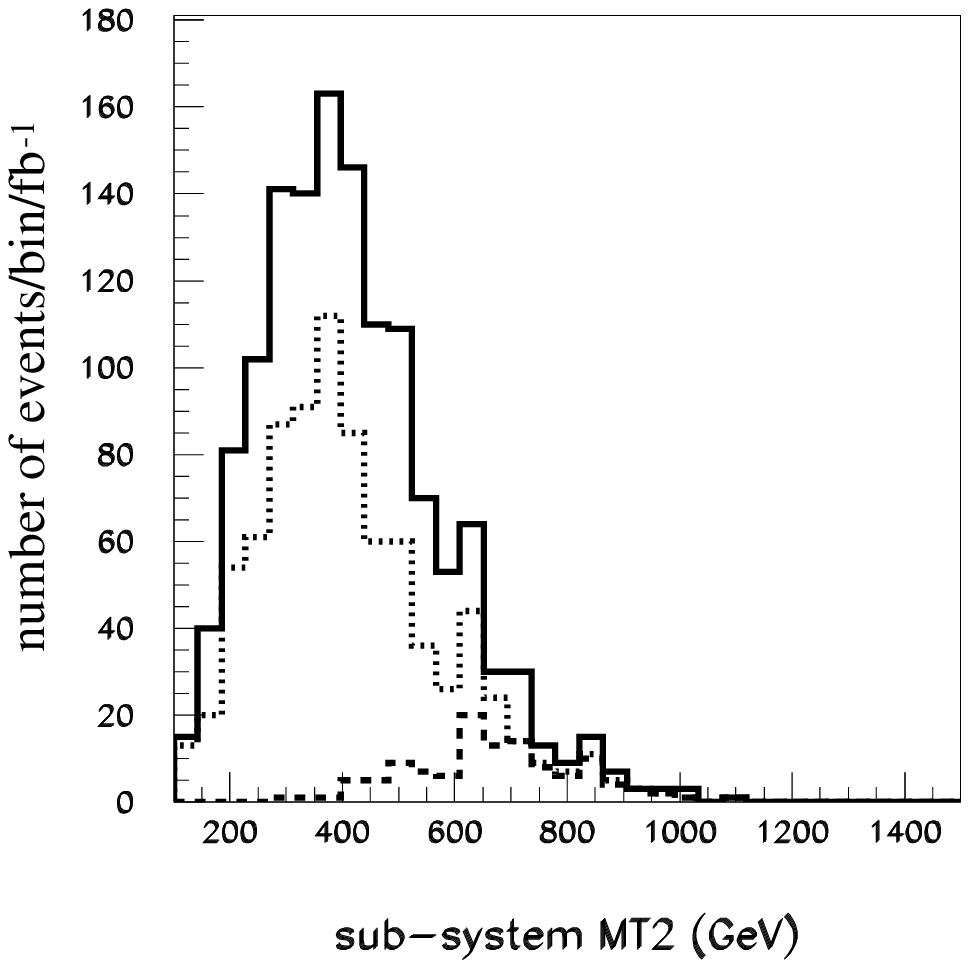}
\caption{The same as Fig.~\ref{Figure:pt1jets} but at point f.}
\label{Figure:pt1jetf}
}

\FIGURE[!ht]{
\includegraphics[width=5cm]{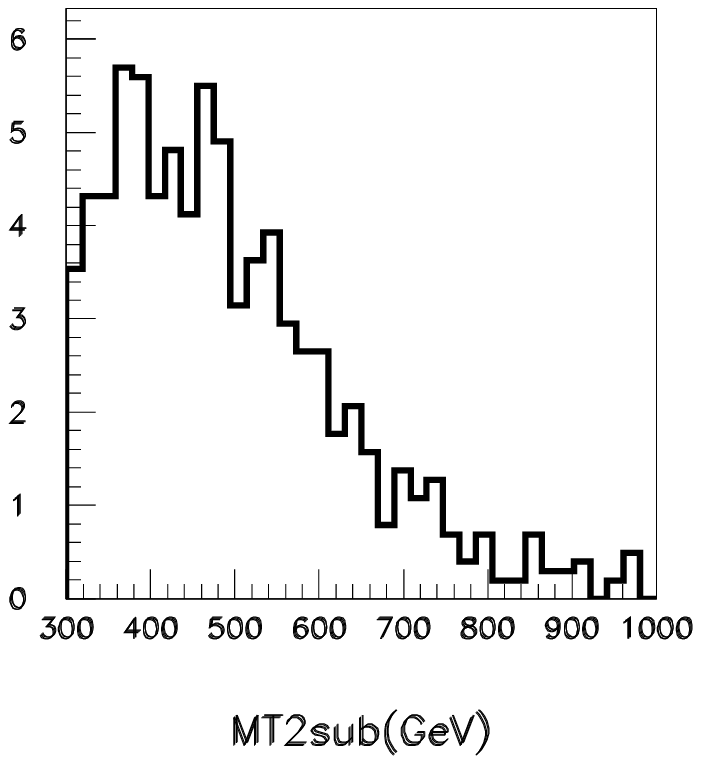}\hskip -1cm
\includegraphics[width=5cm]{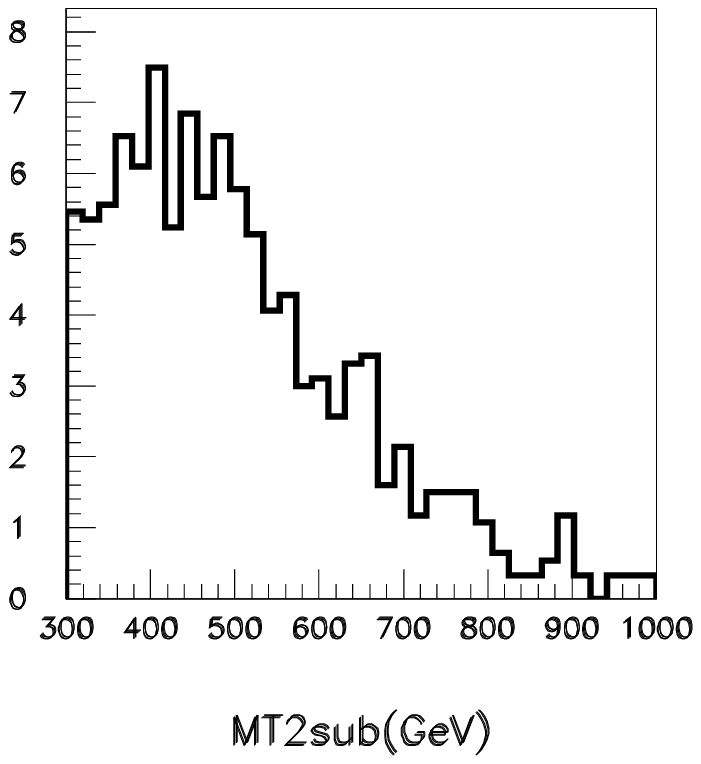}\hskip -1cm
\includegraphics[width=5cm]{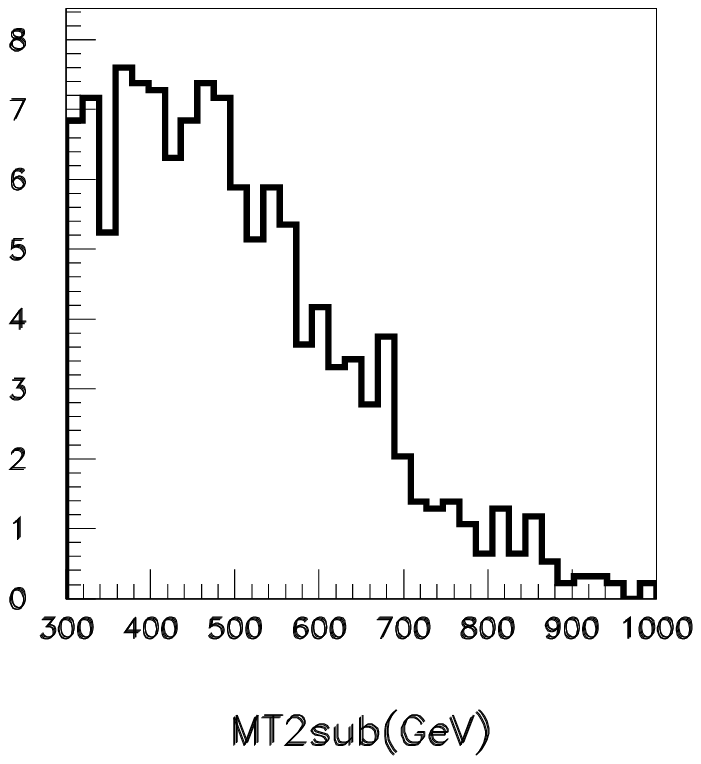}\\
\includegraphics[width=5cm]{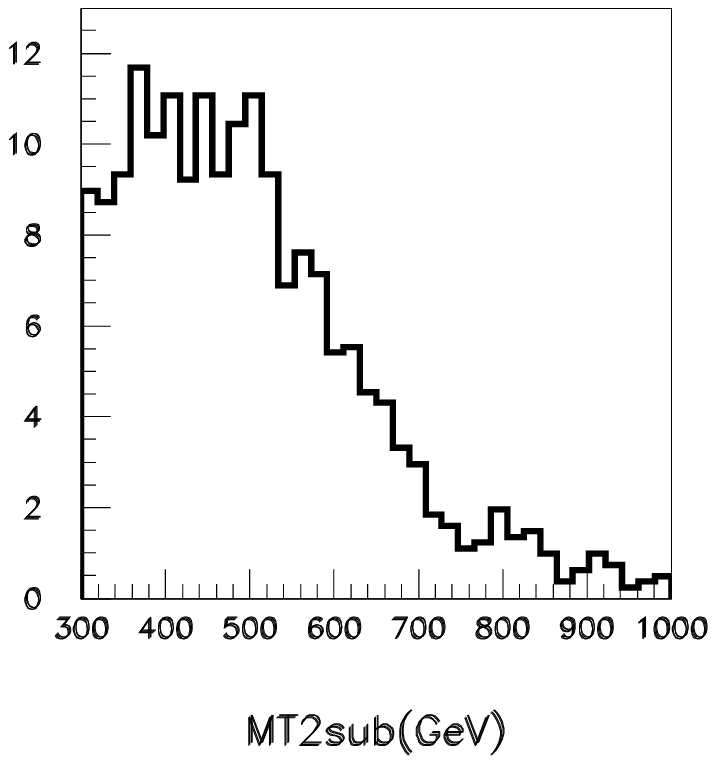}\hskip -1cm
\includegraphics[width=5cm]{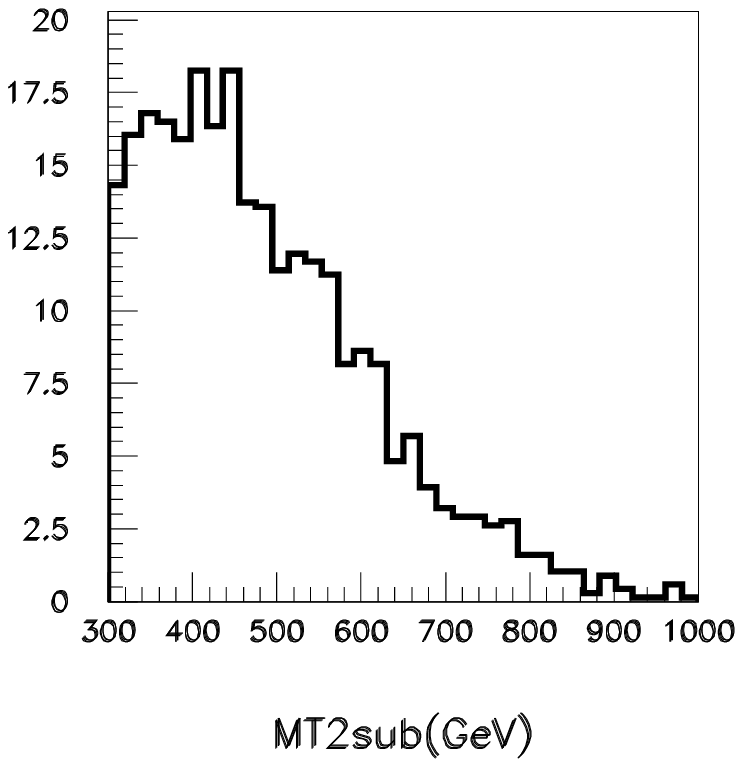}\hskip -1cm
\includegraphics[width=5cm]{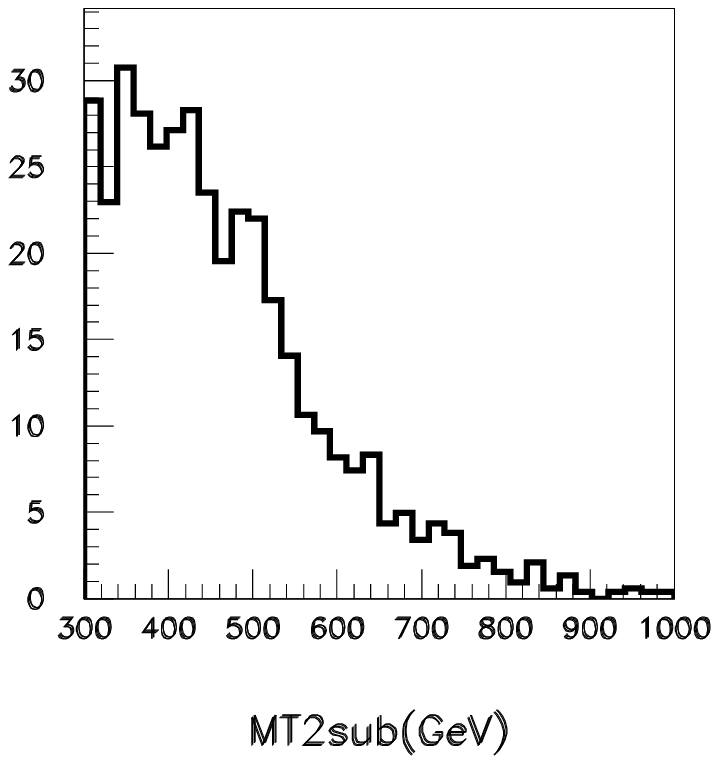}
\caption{ The $m^{\rm sub}_{T2}$ destributions at points a to f
normalized for $1\,$fb$^{-1}$. The input gluino masses differ slightly 
among the model points as we fix gaugino mass at the GUT scale.  
The top figures  correspond to points a, b and 
c from  left to right, and  $m_{\tilde{g}}= 
 796, 785$, and 780~GeV respectively. The bottom figures correspond 
 to  points d, e and  f,  and $m_{\tilde{g}}=
 773, 762$ and 745~GeV.  The squark mass is 1516~GeV at point a and 
  881~GeV   
 at point f. 
  }\label{Figure:rmt2subs_1fb}
}

We now demonstrate   
the gluino mass determination using  the endpoint of the 
$m^{\rm sub}_{T2}$ distribution.   Here we must pay 
some attention to reduce the contributions from the squark-squark pair productions,
which give the endpoints of the $m^{\rm sub}_{T2}$ distribution as $m_{\tilde{q}}$.
This contribution smears the 
endpoint at the gluino mass. 
It is important to find the cuts to reduce the events.

We find that the cut on the number of high $p_T$ jets above a certain threshold 
is useful to reduce the contamination, 
becuase the squark decay tends to give high $p_T$ jets, 
as we have discussed earlier. 
To see this, we first show the distributions 
of the highest $p_T$ jet at point a for 50,000 generated events in Fig.~\ref{Figure:pt1jets}. 
The solid lines show the $p_T(1st) $ distribution, 
where $p(1st)$ is the momentum of the highest $p_T$ jet 
among the jets with  $\vert \eta \vert <3$. 
The dashed lines show the contribution from  the events with 
$n_{300}\equiv N_{\rm jet}(p_T>300$~GeV$,\vert \eta \vert <3)=1$ and the dotted lines show 
the contributions of the  events  with 
$n_{\tilde{q}}=1$, where $n_{\tilde{q}}$ is 
the number of primary produced 1st generation squarks of the events.  
The standard SUSY cuts 
are applied to the events. We can see that most of the events with 
$p_{T}(1st) >300$~GeV satisfy  $n_{300}=1$ and 
 they  mostly come from squark-gluino productions.  
Therefore, if $n_{300} \ge 2$, they are likely come from squark-squark pair production events.

Based on the above observation, 
we calculate the $m^{\rm sub}_{T2}$ distribution
only for the events 
which have only one or zero high $p_T$ jet above a certain $p_T$ threshold. 
The actual value of the  $p_T$ cut
should be chosen based  on the  signal distribution. 
For our model points, we take the cut  $n_{300}\le 1$. 
We do not include the events 
with $n_{300} \ge 2$,  
because our MC simulations show 
that they mostly  come from the
squark pair production. 
In the right figure, we show  $m_{T2}^{\rm sub}$ distributions  
for the events $n_{300}=1$ at point a. 
The dotted line shows the distribution with $n_{300}=1$ 
and $n_{\tilde{q}}=1$. 
The dashed line is the distribution with   $m_{T2}>800$~GeV, 
$n_{300}=1$ and  $n_{\tilde{q}}=1$.  All distributions show the 
$m_{T2}^{\rm sub}$ endpoint close to the gluino mass value $\sim 800$~GeV,  
which is expected from the parton level analysis. 

Fig.~\ref{Figure:pt1jetf} shows the same distributions at point f.  
Events from  squark-gluino co-production still 
dominate the events with $p_T>300$~GeV, 
and a significant fraction of the events satisfy $n_{300}=1$. 
The events near the $m^{\rm sub}_{T2}$  endpoint 
 mostly come from squark-gluino production.   The endpoint of the 
distribution $\sim 750$~GeV is consistent  with the  gluino mass. 

The $m_{T2}^{\rm sub}$ distributions at points a to f for $1\,$fb$^{-1}$ of integrated luminosity
are shown in Fig.~\ref{Figure:rmt2subs_1fb}.  
Here we require $n_{300}\le 1$; therefore,
the distributions now include significant events from gluino-gluino production 
 unlike the  previous plots.  We have seen that  the  $m_{T2}$ distribution 
 changes   significantly among points a to f.   The  $m_{T2}^{\rm sub}$  
distributions are, by contrast,  similar. This is because  the 
$m^{\rm sub}_{T2}$ endpoints 
must be very close to the true gluino mass  $m_{\tilde{g}}\sim 750$~GeV 
(up to the difference 
of the test LSP mass from the true LSP mass). 
This is also seen in Fig.~\ref{Figure:mt2fit}, 
where the value of the fitted 
$m^{\rm sub}_{T2}$  endpoint is shown  together with 
the gluino mass for each point.

\section{Background $m_{T2}$ and $m^{\rm sub }_{T2}$ 
\label{sec;BG}
distributions} 
The Standard Model background to the SUSY processes has been studied by 
ATLAS and CMS groups extensively. The ratio $E\slush_{T}$ $/M_{\rm eff}$
gives a good discrimination between the SUSY signal and the background. 
In the previous section we required $E\slush_{T}/M_{\rm eff}>0.2$ in 
addition to $M_{\rm eff}> 500$~GeV and $E\slush_{T}>200$~GeV. 

The production cross section of the SM background is huge compared with 
the typical signal cross section. To measure the endpoint of the 
signal $m_{T2}$ and $m^{\rm sub}_{T2}$ distributions, the signal to  
noise  ratio ($S/N$) must be sufficiently small near the endpoint. 
The SM backgrounds in the 0-lepton channel after the standard SUSY cuts come
from the four different sources: $t\bar{t}$, $W^{\pm}$, $Z^0$ productions
 with multiple jets, and QCD multi-jet processes. 
Bottom quark productions and 
the  mis-measurements of particle energies can give the missing energy 
to QCD multi-jet processes. 
It is difficult to estimate the QCD background without knowing detector 
performances in detail. We therefore do not attempt to do so in this paper.  
In recent ATLAS and CMS studies \cite{LHCsusy08}, the four channels
 contribute to the background at roughly the same order of magnitude 
after the cuts to reduce the SM backgrounds, 
although QCD background decreases  much faster with increasing  $M_{\rm eff}$.

\TABLE[!ht]{
\begin{tabular}{|l || r r r r |}  
 \hline
                                          &  $t\bar{t}$ & $W$ &$Z$ & total    \cr     
\hline
$m_{T2}>500$~GeV         &     77.7  &   104.9     &   107.0    &   289.6   \cr
                                          &   38.4    & 44.8 & 39.9 & 123.1 \cr
\hline
$m_{T2}>700$~GeV         &      20.3   &    24.4  &     23.4  &      68.2   \cr
                                          &      10.0  &  12.0 & 10.4   & 32.4 \cr
\hline
$m_{T2}^{\rm sub}>300$~GeV   &       90.3       &  80.4      &     82.2      &  252.9      \cr
                                          &       44.2 &  38.5& 31.7 &  113.1 \cr
$m_{T2}^{\rm sub}>500$~GeV  &        11.1 &  6.9 & 6.1 &  24.0 \cr
                                          &       8.1    &  4.9      &  3.8         &  16.8\cr
\hline
luminosity          &    13.1 fb$^{-1}$  & 13.5 fb$^{-1}$&  19.1 fb$^{-1}$&
\cr
\hline
\end{tabular}
\label{Table:background}
\caption{Number of SM background events per $1\,$fb$^{-1}$. 
For each row, upper (lower) numbers 
correspond to the events without (with) a cut on the hemisphere masses, 
$m_{\rm hemi}>200$~GeV. 
The last row 
shows the number of generated events for this study in terms of 
the corresponding integrated luminosity.}
}

The source of missing $E_T$ for the processes $t\bar{t}$, $Z^0$, and $W^{\pm}+n$ jets 
is primarily escaping neutrinos, and missing $E_T$ arising 
from energy  mis-measurements is less important.
We generate these events using {\tt ALPGEN} \cite{Mangano:2002ea,Mangano:2001xp}, and parton shower and initial 
state radiations are estimated by interfacing the parton level events 
to {\tt HERWIG}.   
We generate 
 $Z^{0}(\rightarrow \nu \bar\nu) +n$ jets for $n\le 5$ , 
$W^{\pm}(\rightarrow l\nu)+n$ jets ($n\le 4$) , and   $t\bar{t}+n$ jets ($n\le 2$) , 
so that tree level 0 lepton events have at least 4 or 5 jets 
including $\tau$ jets.
We  require minimum parton separation $\Delta R_{jj}>0.6$\footnote{
The jet cone size for  {\tt AcerDET} jet reconstruction  is set to $R=0.4$. This 
means that centers of two well separated jets has $\Delta R> 0.8$. 
We therefore require $\Delta R_{jj}$ to be slightly lower than that. This 
is sufficient for our purpose as we are working on 
inclusive signatures.   
Reducing the $\Delta R_{jj}$ cut to less than 0.6 
results in unnecessary inefficiency to the event 
generation.}, 
and place a cut on the 
forward parton of $\vert \eta\vert <5$.  
The events are then matched so that there 
is no double counting between parton shower and hard partons 
by using the MLM matching scheme provided by {\tt ALPGEN}.
In this scheme, we generate the processes with up to $n_{max}$ parton.
The events from the processes with $n$ partons ($n<n_{\rm max}$) 
are accepted only if jets and partons match ($n_{\rm jet} = n$),
while the events from the processes with $n_{\rm max}$ partons
are accepted if $n_{\rm jet} \ge n$.
In order to reduce the number of produced events while keeping 
enough statistics for the kinematical region we are interested in, 
we require $\sum_{\rm parton} E_T > 400$~GeV  for $W$+ $n$ jets,
$E\slush_{T}> 150$~GeV for $Z+ n$ jets, 
$\sum_{\rm parton} E_T> 500$~GeV for $t\bar{t}$+ $n$ jets\footnote{
The conditions 
of the generations for the different processes are not the same. 
However, these conditions are loose enough 
so that there is  no effect of the generation cuts 
after our standard SUSY cuts.}.
The effect of additional jets on the signal distributions is small
and discussed in Appendix \ref{sec;app_ME}.

{\tt AcerDET} performs Gaussian smearing for jet momenta, 
the missing momentum,  and isolated lepton momenta.  
It does not contain various potentially important instrumental effects, such as 
non-Gaussian tails of the energy smearing and  lepton inefficiencies. 
Therefore,  our background estimate is given in this paper 
for  illustrative purpose, and more realistic estimates must be performed  by the experimental groups. 

Keeping this in mind, 
Table \ref{Table:background} summarizes results of our event generations. 
The number of background events for $\int dt {\cal L}=1\,$fb$^{-1}$ under various cuts 
are given. 
The bottom row shows  the  corresponding luminosities 
we have generated for the background processes.  
We apply the SUSY cuts given  in Section 4. 
In addition,  
we require $n_{300}\le 1$ for $m_{T2}^{\rm sub}$  distribution. 
We do not include $K$ factors, as the corresponding higher order 
QCD corrections are  not available. 
Note that K factors of 
$t\bar{t}$ production and SUSY production tend to 
cancel partially.  
For each row, upper (lower) numbers 
correspond to the events without (with) a cut on the hemisphere masses, 
$m_{\rm hemi}>200$~GeV. 
The background with the hemisphere mass  cut
is reduced by more than a factor of 2. 
This suggests that the background events are dominated by
the configurations
that a few jets are either soft or colinear to leading hard jets
and therefore the masses of the hemispheres are small.
The background distributions will be studied in detail elsewhere. 

\FIGURE[!ht]{
\includegraphics[width=6cm]{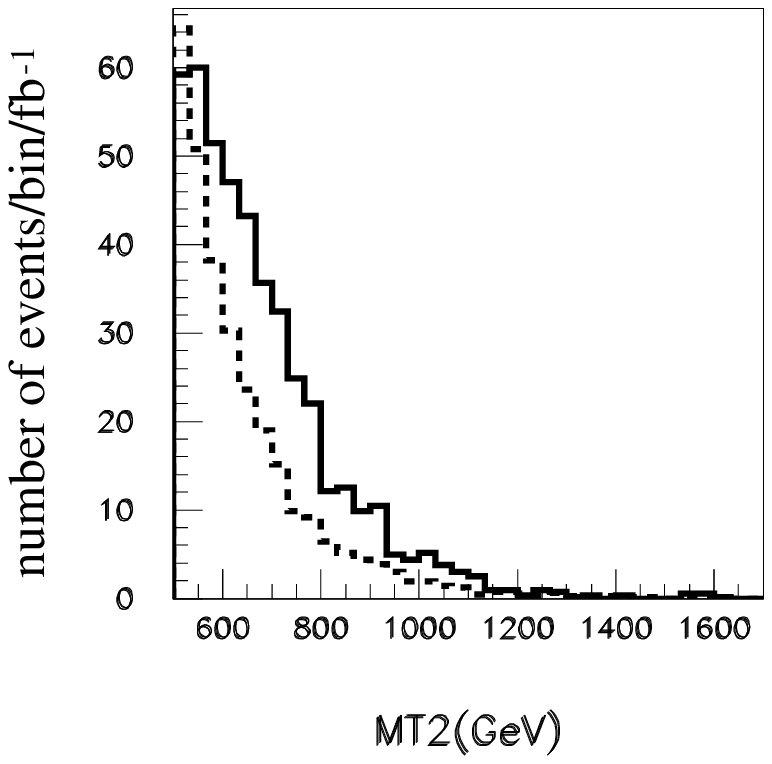}\hskip -1cm
\includegraphics[width=6cm]{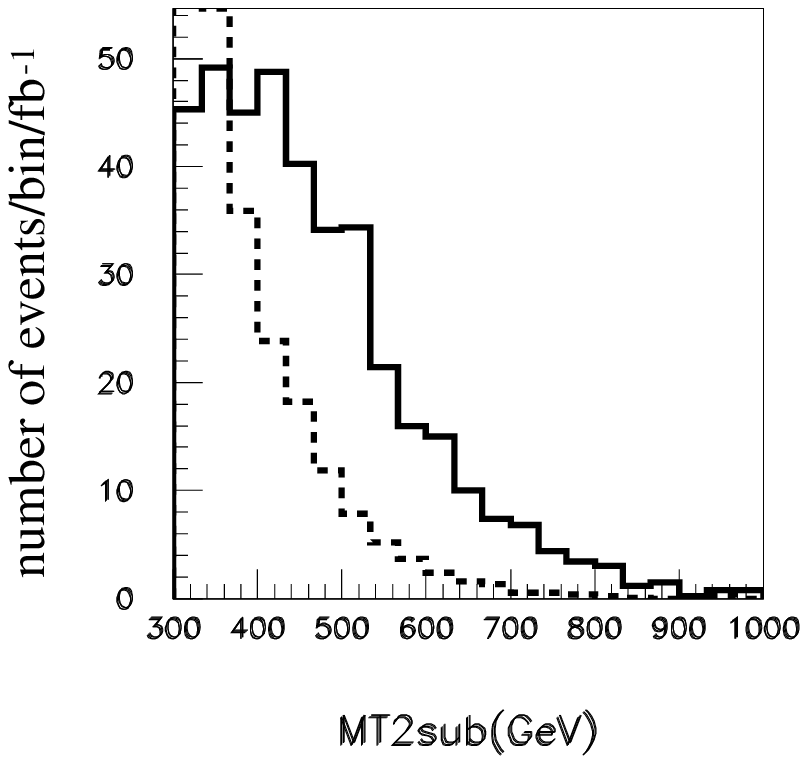}\hskip -1cm
\includegraphics[width=6cm]{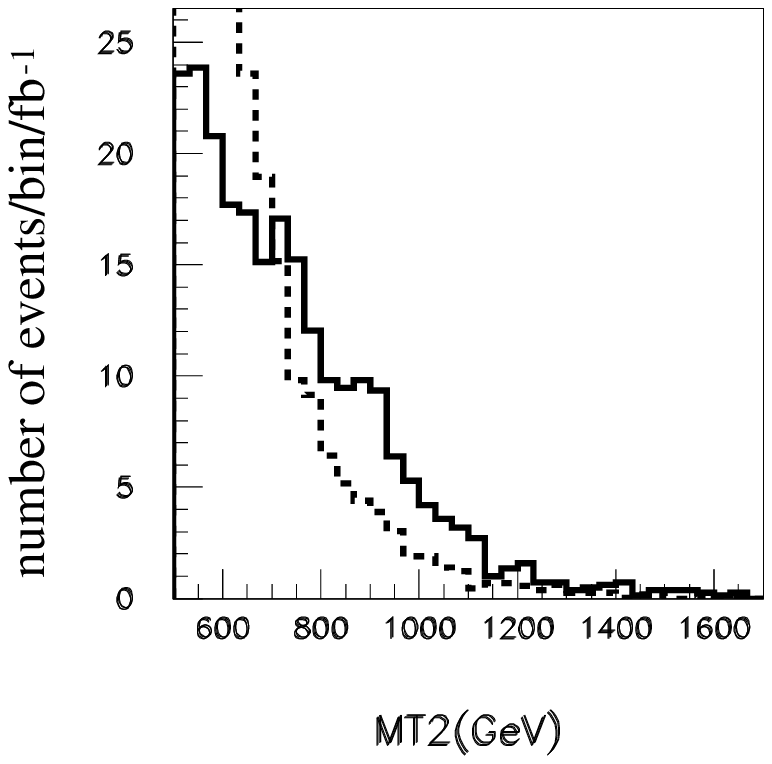}\hskip -1cm
\includegraphics[width=6cm]{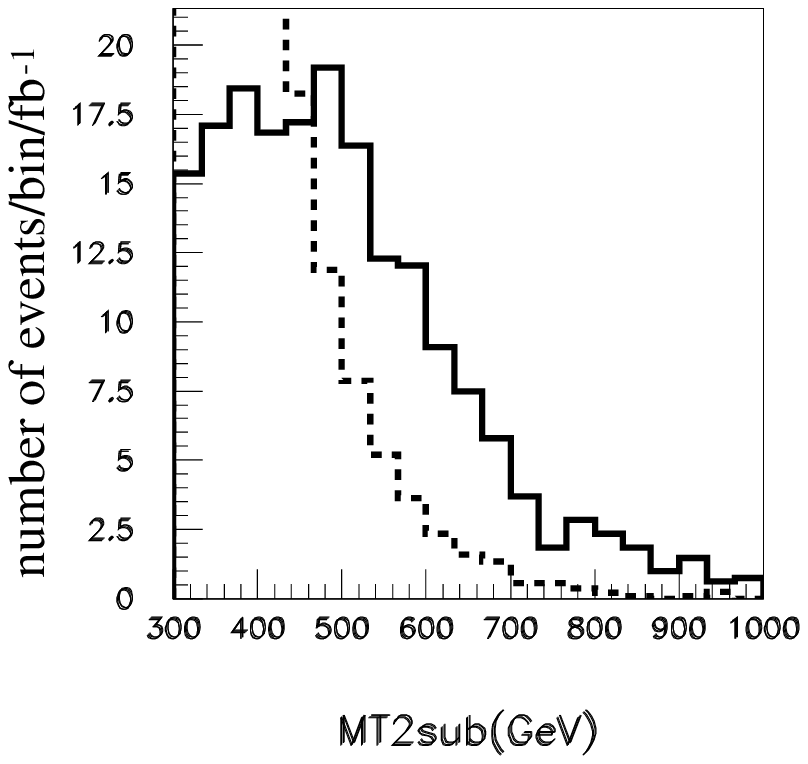}\hskip -1cm
\caption{Signal and background distributions at point f (the top figures) and 
at point d (the bottom figures).
Unit of $y$-axes is events/bin/$1\,$fb$^{-1}$.
}\label{Figure:sgbgfd}
}
\FIGURE[!ht]{
\includegraphics[width=6.cm]{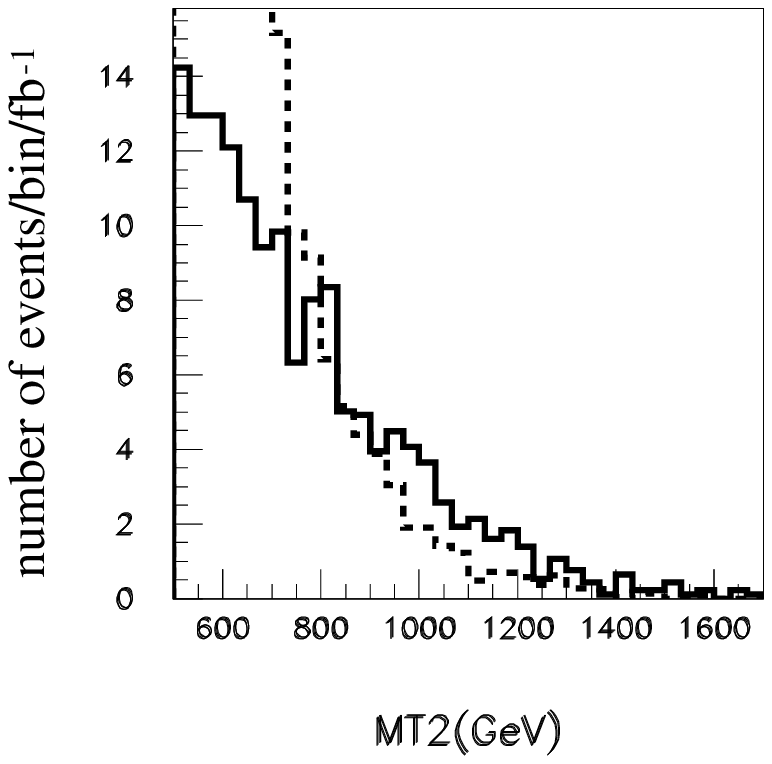}\hskip -1cm
\includegraphics[width=6.cm]{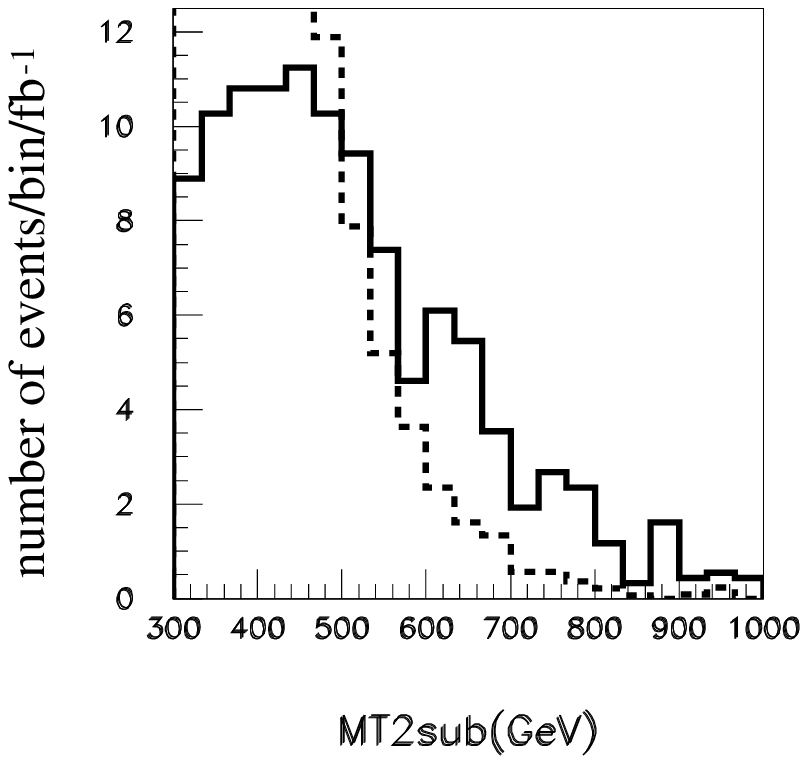}\hskip -1cm\\
\includegraphics[width=6.cm]{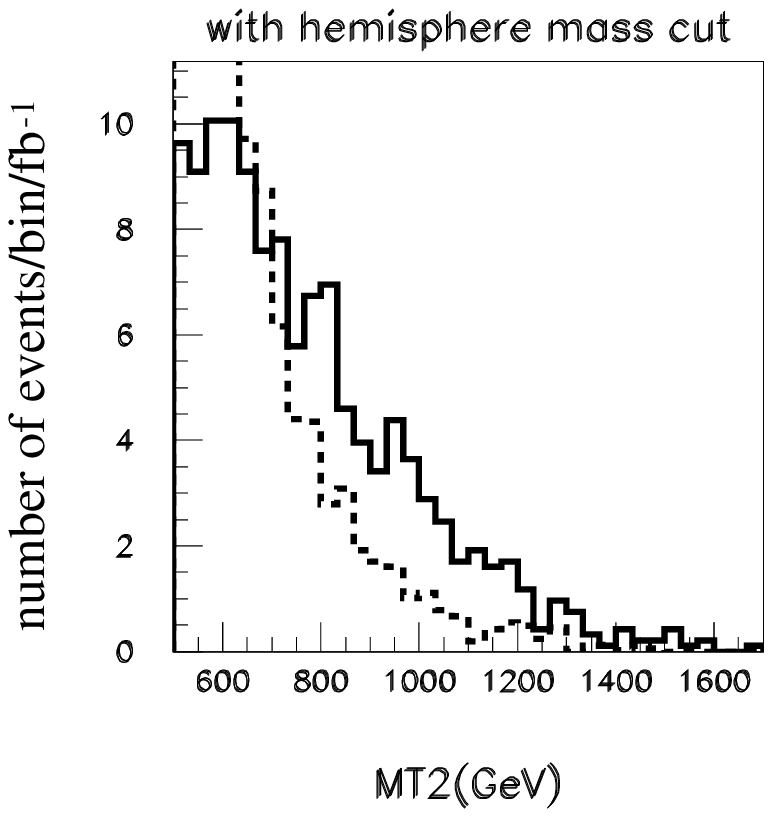}\hskip -1cm
\includegraphics[width=6.cm]{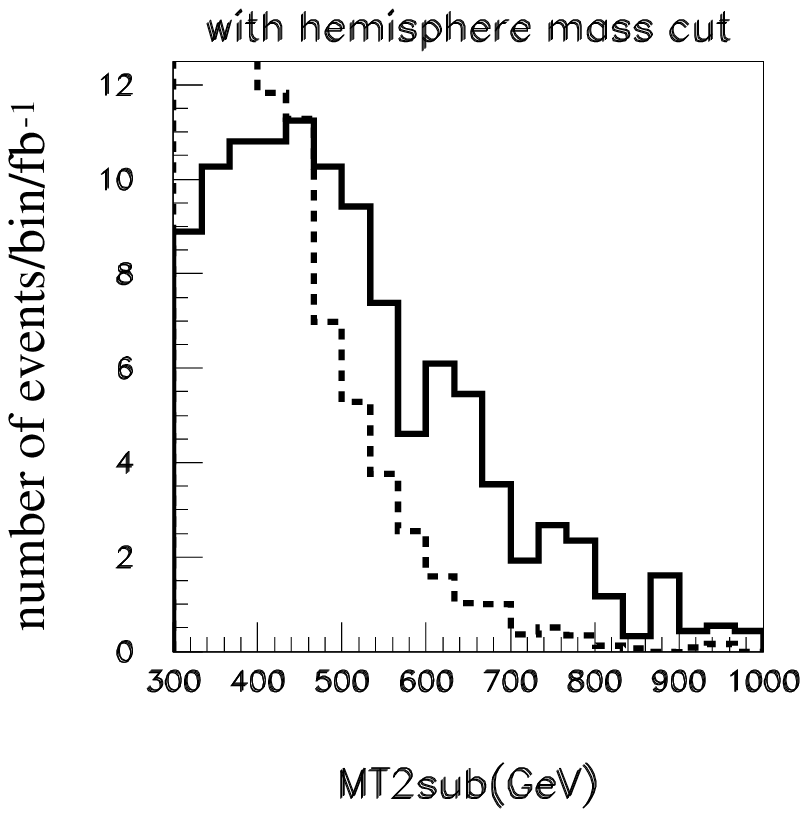}
\caption{The signal (solid) and background (dashed) 
 distributions at point b.  The bottom figures 
show the distributions with the hemisphere mass cut $m^{1(2)}_{\rm hemi}>200$~GeV. 
Unit of $y$-axes is events/bin/$1\,$fb$^{-1}$.
 }
\label{Figure:sgbgb}
}

Fig.~\ref{Figure:sgbgfd} shows the distribution of  background, together with the 
signal distribution at points f (the top figures) and d (the bottom figures). 
These distributions are without 
hemisphere mass cuts. 
The signal is larger than the background above $m_{T2}>600$ $(750)$~GeV at points f (d) 
for the  $m_{T2}$ distribution, which is much smaller than expected 
$m_{T2}^{\rm end} =m_{\tilde{q}} =881$~GeV ( $1175$~GeV).
The endpoints of the signal $m_{T2}$ and $m^{\rm sub}_{T2}$ distributions 
may be extracted as a kink in the total distribution in this case. 
The signal and background distributions of $m_{T2}^{\rm sub}$ 
are also shown in the right plots. 
Again, the level of the background is small near the endpoint. 

We also show the same distribution  at point  b in Fig.~\ref{Figure:sgbgb}.  
The signal cross section involving $\tilde{q}$ production 
is reduced by a factor of $1/5$ from that at point f (See Table \ref{modelbr}).
The $S/N$ above $m_{T2}>700$~GeV is now $\sim 1$
and the cross point of the signal and the background is at $m_{T2}=800\,$GeV.  
By applying the hemisphere mass 
cut, we can reduce the background significantly. 
The improvement of $S/N$ 
near the endpoint can be seen by comparing the top and 
bottom figures without/with  the hemisphere mass cut. 
It is important to reduce the background to measure 
the squark and gluino masses near the discovery regions. 

\section{Conclusions}
\label{sec;conclusions}
The ATLAS and CMS experiments at the LHC  can discover squarks and gluinos in the MSSM with masses
less than 1.5 TeV 
at the early stage of the experiment  with luminosity 
around $\int dt {\cal L}=1\,$fb$^{-1}$.
Developing a reliable method of estimating 
squark and gluino masses with the discovery is 
an important step to study supersymmetry at the LHC.

For this purpose we 
cannot rely on the clean golden channels such as 
$l^\pm l^\mp +$ jets, becuase they tend to have small branching 
ratios and are sensitive to the model parameters. 
In a previous paper\cite{Nojiri:2008hy},
we defined an inclusive $m_{T2}$ variable.
This variable can be calculated for any event with jets and missing transverse 
energy.  It is  calculated in two steps; 
we first define 
the two hemisphere axes by assigning  particles into the two leading 
jets of the events, 
then, 
the $m_{T2}$ variable is caluculated from the two hemisphere 
momenta and missing transverse energy.  We pointed out that 
the endpoint of the 
$m_{T2}$ distribution is sensitive to the squark mass for the case 
$m_{\tilde{q}}>m_{\tilde{g}}$. 

In this paper, we define a \lq\lq sub-system'' $m_{T2}$, $m^{\rm sub}_{T2}$. 
This is an $m_{T2}$ variable calculated without including 
the highest $p_T$ jets for the hemisphere assignments and $m_{T2}$ calculation.  
In the case that
 $m_{\tilde{q}}>m_{\tilde{g}}$ and  the other sparticles are lighter, 
the endpoint of $m^{\rm sub}_{T2}$ distribution gives us information on $m_{\tilde{q}}$.  
In this paper, we show convincing evidence for sample model 
 points within the reach for $\int dt L=1\,$fb$^{-1}$. 
 
We also provide various parton level checks on the hemisphere algorithm. 
We estimate background 
distributions arising from $t\bar{t}$+$n$ jets, $Z^0$+ $n$ jets  and
$W^{\pm}$ + $n$ jets using {\tt ALPGEN} and find out that $S/N$ ratio is 
large for the events near the $m_{T2}$ endpoints at our sample points.
In the Appendix, 
we also provide a study of SUSY$+ n$ jet distributions using {\tt MadGraph}{\tt /MadEvent}, 
and find that the endpoint is stable with the ME corrections.   

\acknowledgments
We would like to thank to Rikkert Frederix for help with using Madgraph
and to Willie Klemm for careful reading of the manuscript.
This work is supported in part by World Premier International
    Research Center InitiativeiWPI Initiative), MEXT, Japan,. M.M.N. and K.S. are 
    supported  in part by 
     the Grant-in-Aid for Science Research,MEXT, 
Japan .

\appendix
\section{Appendix}
\subsection{The $m_{T2}$ endpoint for squark-gluino production events}
\label{sec;app_mt2}
In this Appendix, we show the condition for which
the endpoint of the ideal $m_{T2}$ distribution for the squark-gluino production events
coincides with the squark mass at $M_{\rm test}=m_{\rm LSP}$.

The squark-gluino $m_{T2}$ is calculated by minimizing 
$\max \{m_{T}^{(\tilde{q})}, m_{T}^{(\tilde{g})} \}$
under the condition that the sum of transverse test momenta of two LSP
is equal to the $E\slush_T$.
It is known that the transverse mass $m_{T}^{(i)}$ ($i=\tilde{q},\tilde{g}$) 
as a function of the test LSP momentum has the global minimum,
which is called the {\it unconstrained minimum} (UCM) \cite{Barr:2003rg}.  
There are cases where $m_{T2}$ is given by the {\it unconstrained minimum} 
of the transverse mass on one side $(m_{T}^{(i)})_{\rm UCM}$.
This situation occurs when 
$m_T^{(j)}$ on the other side for the test LSP momentum which gives the $(m_T^{(i)})_{\rm UCM}$
is smaller than $(m_T^{(i)})_{\rm UCM}$.

In Ref.\cite{Cho:2007qv}, it is shown that 
the UCM of the squark system ($(m_T^{(\tilde{q})})_{\rm UCM}$) is given by
\begin{equation}
(m^{(\tilde{q})}_{T})_{\rm UCM} = m^{(\tilde{q})}_{\rm vis} + M_{\rm test},
\label{ucmdef}
\end{equation}
where $m^{(\tilde{q})}_{\rm vis}$ is the invariant mass of the visible particles from the squark decay.
The maximum of $(m_T^{(\tilde{q})})_{\rm UCM}$ is, therefore,
given by substituting the maximum of $m^{(\tilde{q})}_{\rm vis}$ into Eq.\ref{ucmdef}.  
The maximum of the $m^{(\tilde{q})}_{\rm vis}$ is given by
\begin{equation}
(m^{(\tilde{q})}_{\rm vis})^{\rm max} = m_{\tilde{q}} - m_{\rm LSP},
\end{equation}
if the LSP from squark decay can be at rest in the squark rest frame.
In this case, the maximum of the UCM of the squark system 
can reach the squark mass at $M_{\rm test}=m_{\rm LSP}$,
and the maximum of the squark-gluino $m_{T2}$ is identical to the squark mass at $M_{\rm test}=m_{\rm LSP}$. 

\FIGURE
{
\includegraphics[width=7cm]{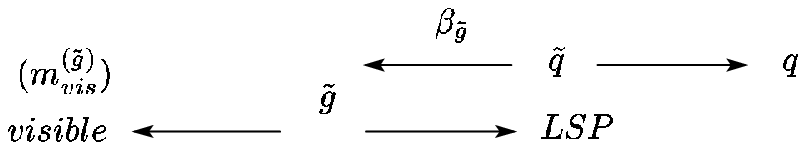}
\caption{A Kinematical configuration of squark decay}\label{Figure:sqdecay}
}

We now consider the condition that the LSP can be at rest in the squark rest frame.
In the following discussion, we concentrate on the case 
that the squark decays into the gluino and a jet.
The gluino from the squark subsequently decays into 
the visible objects and the LSP (See Fig.~\ref{Figure:sqdecay}).
The LSP momentum in the gluino rest frame ($p'_{\rm LSP}$) 
depends on the invariant mass of the visible objects ($m^{(\tilde{g})}_{\rm vis}$) as
\begin{equation}
|p'_{\rm LSP}|=\frac{1}{2 m_{\tilde{g}}} \sqrt{
m_{\tilde{g}}^4 -2m^2_{\tilde{g}} ((m^{(\tilde{g})}_{\rm vis})^2 + m^2_{\rm LSP}) 
+((m^{(\tilde{g})}_{\rm vis})^2-m^2_{\rm LSP})^2
}~~.
\end{equation} 
If the LSP is produced in the opposite direction from the gluino momentum
and the gluino velocity is not too large,
 the LSP can be at rest in the squark rest frame for a suitable value of the invariant mass 
of the visible objects ($\tilde{m}_{\rm vis}^{(\tilde{g})}$).
In this situation,
the LSP momentum in the squark rest frame ($p_{\rm LSP}$) is obtained 
by the Lorentz boost of the $p'_{\rm LSP}$ as
\begin{equation}
p_{\rm LSP}=\gamma_{\tilde{g}} (-|\beta_{\tilde{g}}| E'_{\rm LSP} + |p'_{\rm LSP}|)~,
\end{equation} 
where $E'_{\rm LSP}$ is the energy of the LSP in the gluino rest frame,
and the Lorentz boost factors $\beta_{\tilde{g}}$ and $\gamma_{\tilde{g}}$ are given by
\begin{equation}
|\beta_{\tilde{g}}|=\frac{m^2_{\tilde{q}}-m^2_{\tilde{g}}}{m^2_{\tilde{q}}+m^2_{\tilde{g}}},
~~~~~
\gamma_{\tilde{g}} = 1/\sqrt{1-\beta^2_{\tilde{g}}} 
=\frac{m^2_{\tilde{q}}+m^2_{\tilde{g}}}{2 m_{\tilde{q}}m_{\tilde{g}}}.
\end{equation} 
By solving the equation $p_{\rm LSP}=0$, we obtain
\begin{equation}
(\tilde{m}_{\rm vis}^{(\tilde{g})})^2 = 
m_{\tilde{g}}^2 \big( 1-\frac{m_{\rm LSP}}{m_{\tilde{q}}} \big) 
\big( 1-\frac{m_{\tilde{q}} m_{\rm LSP}}{m^2_{\tilde{g}}} \big)~.
\label{mvg_sol}
\end{equation}

Note that if $m_{\tilde{q}} m_{\rm LSP} > m^2_{\tilde{g}}$, 
the equation $p_{\rm LSP}=0$ does not have any solution 
for positive $\tilde{m}_{\rm vis}^{(\tilde{g})}$.
In this case, the LSP cannot be at rest in the squark rest frame, 
and $(m^{(\tilde{q})}_{T})_{\rm UCM}$ is less than the squark mass.
Even if the equation $p_{\rm LSP}=0$ has a solution for positive $\tilde{m}_{\rm vis}^{(\tilde{g})}$,
there are cases where $m_{\rm vis}^{(\tilde{g})}$ has a non-vanishing kinematical lower bound
due to a heavy standard model particle, such as $t$, $Z$ and $W$.
If the lower bound is smaller than the solution $\tilde{m}_{\rm vis}^{(\tilde{g})}$,
$(m^{(\tilde{q})}_{T})_{\rm UCM}$ cannot reach the squark mass.
For our model points, the solution (\ref{mvg_sol}) is
$\tilde{m}_{\rm vis}^{(\tilde{g})}/m_{\tilde{g}} = 0.83$ for point a,
 and $\tilde{m}_{\rm vis}^{(\tilde{g})}/m_{\tilde{g}} = 0.85$ for point f.
On the other hand, the kinematically allowed range of the visible invariant mass is roughly
$0.50 \le m_{\rm vis}^{(\tilde{g})}/m_{\tilde{g}} \le 0.86$ for point a, and
$0.55 \le m_{\rm vis}^{(\tilde{g})}/m_{\tilde{g}} \le 0.86$ for point f.
Therefore, the endpoint of the ideal $m_{T2}$ distribution in the squark-gluino production events
is identical to the squark mass in our model points.

\subsection{The effect of Matrix Element corrections to the signal distribution}
\label{sec;app_ME}
In this appendix we consider the matching effect of multi-jet matrix 
elements (ME) and  parton showers on the $m_{T2}$ distribution. 
When we calculate the signal $m_{T2}$ distributions in this text, 
we generate SUSY processes at the lowest-order hard process
and then generate multi-jet events by parton showers.
In general, there are ME corrections from hard parton 
emissions in the lowest order hard process, which may not
be included in the parton shower approach. 
Note that we have applied  cuts $p_T> 50\,$GeV and $|\eta|<3$ for the jets 
to be included in the hemispheres.
We need to check that this is enough to kill the effects of initial state radiations.

When the ME corrections are taken into account, we should avoid
double counting of emissions in overlapping phase space
and need some kind of matching scheme to merge the ME corrections. 
Here, we study the ME corrections using the {\tt MadGraph} {\tt /MadEvent} MC generator \cite{Alwall:2007st}, 
in which the matching between the ME corrections and the parton showers 
is implemented. 
For our analysis, we use a modified MLM matching procedure 
with $k_\perp$ jets. In this scheme, the parton emissions are separated
into two phase space regions at some $k_\perp$.
In {\tt MadGraph}{\tt /MadEvent}, only events with enough separated partons,
$k_\perp>${ xqcut}, are generated after the matrix element simulation.
Then parton showering is performed and the partons are clustered
into jets using the $k_\perp$ algorithm. After this procedure,
the matching between the jets and the partons from the matrix elements is performed 
using Pythia. If the distance between them is larger than { Qcut},
the event is discarded in order to avoid double counting.

In order to see the effect of the additional jet emission,
we generate the SUSY events for the mSUGRA point SPS 1a using {\tt MadGraph}{\tt /MadEvent}. 
The generated parton level events are interfaced 
with Pythia to take into account the matching and the hadronization.
We take the matching parameters as
{xqcut}$=40~{\rm GeV}$, {Qcut}$=60~{\rm GeV}.$ 
After hadronic events are generated, we use {\tt AcerDET} for detector simulations.  
We apply the same cuts given in the Sec. 4 to select the events.

In Fig.~\ref{Figure:mt2matching}(a), we plot the $m_{T2}$ distributions
for gluino pair-production processes with 0, 1, 2 jets. 
Since the total cross sections could receive large NLO corrections, 
the shape of $m_{T2}$ distribution is more important. 
For comparison, we normalize the each distribution to unity. 
We can see that the shapes of the $m_{T2}$  distributions 
are stable against the ME corrections.
This is a good feature to obtain information on the gluino mass 
from the endpoint of the $m_{T2}$ distributions.

In Fig.~\ref{Figure:mt2matching}(b) and 12(c), we plot the $m_{T2}$ distributions
for squark pair-production (squark-gluino) processes with 0, 1 jet. 
We also normalize each distribution to unity. 
Again we can see that the $m_{T2}$  distributions are rather insensitive to 
the ME corrections and the matching.

\FIGURE{
\includegraphics[width=4.9cm]{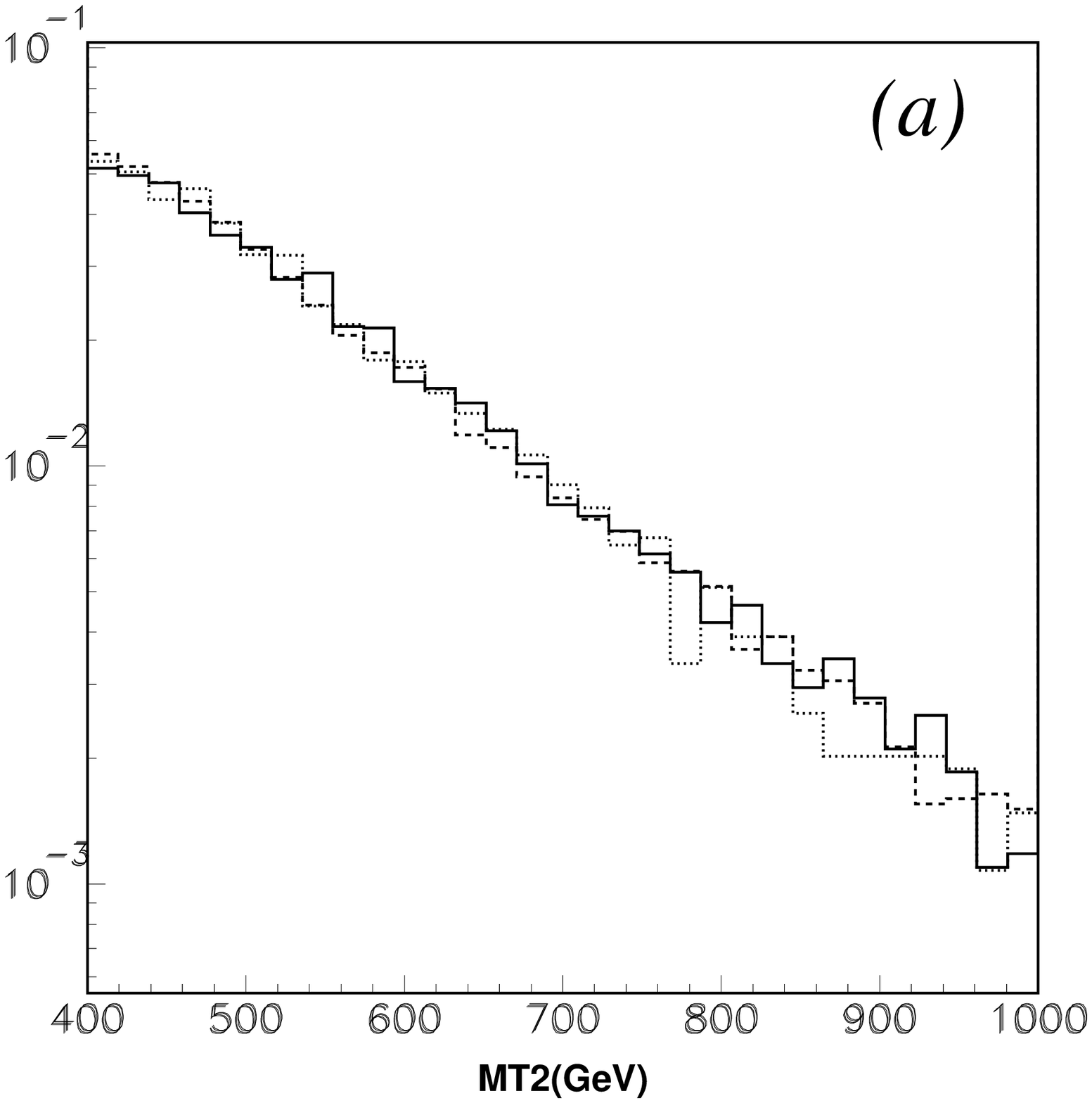}\hfill
\includegraphics[width=4.9cm]{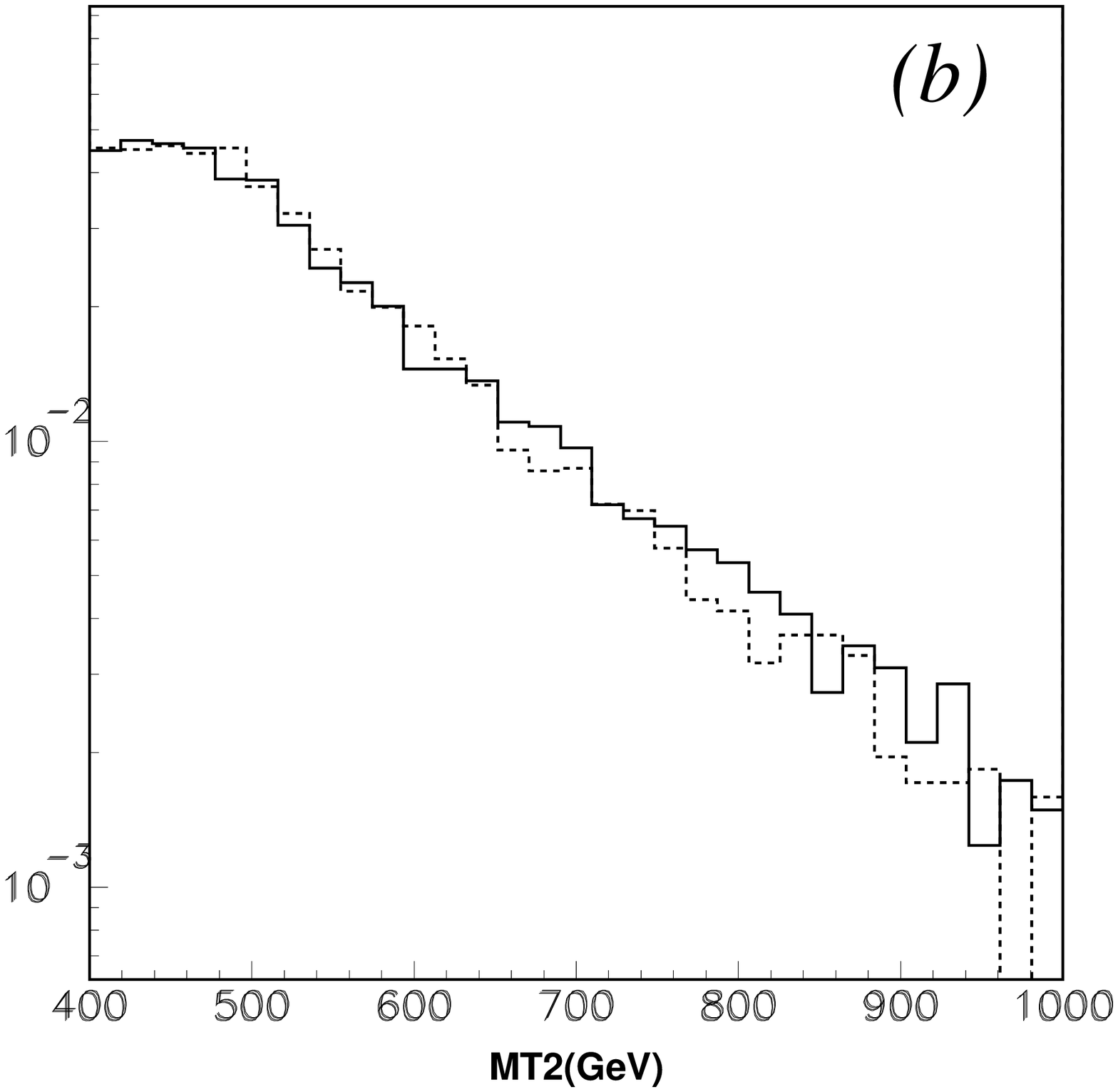}\hfill
\includegraphics[width=4.9cm]{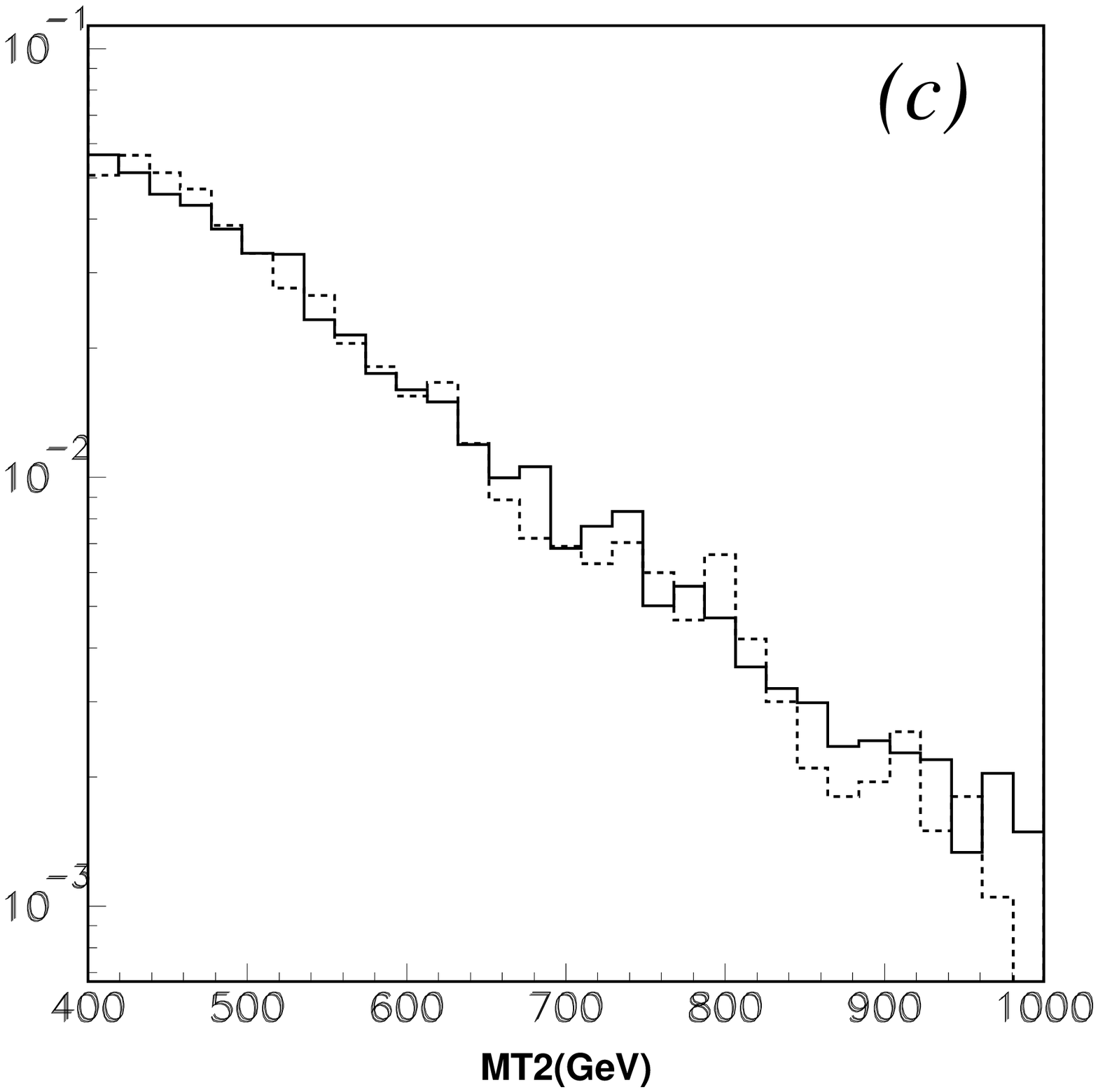}
\caption{(a)  $m_{T2}$ distributions for ${\tilde g}{\tilde g}$ +0 jet (dashed), 1 jet (solid) 2 jets (dotted).
(b)  $m_{T2}$ distributions for ${\tilde q}{\tilde q}$ +0 jet (solid), 1 jet (dashed). (c)  $m_{T2}$ distributions 
for ${\tilde g}{\tilde q}$ +0 jet (solid), 1 jet (dashed). Here SUSY spectrum is assumed as SPS 1a.  } 
\label{Figure:mt2matching}}

        


\begin{thebibliography}{99}

\bibitem{Nilles:1983ge}
  H.~P.~Nilles,
  Phys.\ Rept.\  {\bf 110} (1984) 1.

\bibitem{Haber:1984rc}
  H.~E.~Haber and G.~L.~Kane,
  Phys.\ Rept.\  {\bf 117} (1985) 75.

\bibitem{Martin:1997ns}
  S.~P.~Martin,
  arXiv:hep-ph/9709356.


\bibitem{Hinchliffe:1996iu}
  I.~Hinchliffe, F.~E.~Paige, M.~D.~Shapiro, J.~Soderqvist and W.~Yao,
  Phys.\ Rev.\  D {\bf 55} (1997) 5520
  [arXiv:hep-ph/9610544].


\bibitem{Abdullin:1998pm}
  S.~Abdullin {\it et al.}  [CMS Collaboration],
  J.\ Phys.\ G {\bf 28} (2002) 469
  [arXiv:hep-ph/9806366].


\bibitem{Atlas}
ATLAS Collaboration, 
``ATLAS detector and physic perrformance Technical Design Report,''
CERN/LHCC 99-14/15 (1999).

\bibitem{Bachacou:1999zb}
  H.~Bachacou, I.~Hinchliffe and F.~E.~Paige,
  Phys.\ Rev.\  D {\bf 62} (2000) 015009
  [arXiv:hep-ph/9907518].

\bibitem{Hinchliffe:1999zc}
  I.~Hinchliffe and F.~E.~Paige,
  Phys.\ Rev.\  D {\bf 61} (2000) 095011
  [arXiv:hep-ph/9907519].

\bibitem{Allanach:2000kt}
  B.~C.~Allanach, C.~G.~Lester, M.~A.~Parker and B.~R.~Webber,
  JHEP {\bf 0009} (2000) 004
  [arXiv:hep-ph/0007009].


 
 \bibitem{Nojiri:2003tu}
  M.~M.~Nojiri, G.~Polesello and D.~R.~Tovey,
  arXiv:hep-ph/0312317.
\bibitem{Kawagoe:2004rz}
  K.~Kawagoe, M.~M.~Nojiri and G.~Polesello,
  Phys.\ Rev.\  D {\bf 71}, 035008 (2005)
  [arXiv:hep-ph/0410160].
 \bibitem{Nojiri:2007pq}
  M.~M.~Nojiri, G.~Polesello and D.~R.~Tovey,
  JHEP {\bf 0805}, 014 (2008)
  [arXiv:0712.2718 [hep-ph]].
  
\bibitem{Cheng:2008mg}
  H.~C.~Cheng, D.~Engelhardt, J.~F.~Gunion, Z.~Han and B.~McElrath,
  Phys.\ Rev.\ Lett.\  {\bf 100}, 252001 (2008)
  [arXiv:0802.4290 [hep-ph]].




\bibitem{Lester:1999tx}
  C.~G.~Lester and D.~J.~Summers,
  Phys.\ Lett.\  B {\bf 463} (1999) 99
  [arXiv:hep-ph/9906349].

\bibitem{Barr:2003rg}
  A.~Barr, C.~Lester and P.~Stephens,
  J.\ Phys.\ G {\bf 29} (2003) 2343
  [arXiv:hep-ph/0304226].



\bibitem{Weiglein:2004hn}
   G.~Weiglein {\it et al.}  [LHC/LC Study Group],
   Phys.\ Rept.\  {\bf 426}, 47 (2006)
   [arXiv:hep-ph/0410364].
\bibitem{LHCsusy08}
Oleg Brandt, talk in
Hadron Collider Physics Symposium 2008(HCP)

\bibitem{Cho:2007qv}
  W.~S.~Cho, K.~Choi, Y.~G.~Kim and C.~B.~Park,
  arXiv:0709.0288 [hep-ph]

\bibitem{Gripaios:2007is}
  B.~Gripaios,
  arXiv:0709.2740 [hep-ph].

\bibitem{Barr:2007hy}
  A.~J.~Barr, B.~Gripaios and C.~G.~Lester,
  arXiv:0711.4008 [hep-ph].

\bibitem{Cho:2007dh}
  W.~S.~Cho, K.~Choi, Y.~G.~Kim and C.~B.~Park,
  arXiv:0711.4526 [hep-ph].

\bibitem{Nojiri:2008hy}
  M.~M.~Nojiri, Y.~Shimizu, S.~Okada and K.~Kawagoe,
  JHEP {\bf 0806}, 035 (2008)
  [arXiv:0802.2412 [hep-ph]].


\bibitem{Barr:2008ba}
  A.~J.~Barr, G.~G.~Ross and M.~Serna,
  arXiv:0806.3224 [hep-ph].

\bibitem{Tovey:2008ui}
  D.~R.~Tovey,
  JHEP {\bf 0804}, 034 (2008)
  [arXiv:0802.2879 [hep-ph]].



\bibitem{hemi}
 F.~Moortgat and L.~Pape, CMS Physics TDR, Vol. II, Report
	No. CERN-LHCC-2006, Chap. 13.4, p410

\bibitem{Matsumoto:2006ws}
  S.~Matsumoto, M.~M.~Nojiri and D.~Nomura,
  Phys.\ Rev.\  D {\bf 75} (2007) 055006
  [arXiv:hep-ph/0612249].
\bibitem{Hubisz:2008gg}
  J.~Hubisz, J.~Lykken, M.~Pierini and M.~Spiropulu,
  arXiv:0805.2398 [hep-ph].


\bibitem{Mangano:2002ea}
  M.~L.~Mangano, M.~Moretti, F.~Piccinini, R.~Pittau and A.~D.~Polosa,
  JHEP {\bf 0307}, 001 (2003)
  [arXiv:hep-ph/0206293].

\bibitem{Mangano:2001xp}
   M.~L.~Mangano, M.~Moretti and R.~Pittau,
   Nucl.\ Phys.\  B {\bf 632}, 343 (2002)
   [arXiv:hep-ph/0108069].

\bibitem{Plehn:2005cq}
  T.~Plehn, D.~Rainwater and P.~Skands,
  Phys.\ Lett.\  B {\bf 645}, 217 (2007)
  [arXiv:hep-ph/0510144].
  
\bibitem{talk_alwall}
Johan Alwall,  talk in
"The 16th International Confernce on Supersymmetry and the  
Unification of Fundamental Interactions. "
(SUSY08)

\bibitem{Alwall:2008ve}
  J.~Alwall, M.~P.~Le, M.~Lisanti and J.~G.~Wacker,
  arXiv:0803.0019 [hep-ph].

\bibitem{Alwall:2007st}
  J.~Alwall {\it et al.},
  JHEP {\bf 0709}, 028 (2007)
  [arXiv:0706.2334 [hep-ph]].


\bibitem{Corcella:2000bw}
  G.~Corcella {\it et al.},
  JHEP {\bf 0101} (2001) 010
  [arXiv:hep-ph/0011363];
  arXiv:hep-ph/0210213.

\bibitem{RichterWas:2002ch}
  E.~Richter-Was,
  arXiv:hep-ph/0207355.


\bibitem{Drees:2000he}
  M.~Drees, Y.~G.~Kim, M.~M.~Nojiri, D.~Toya, K.~Hasuko and T.~Kobayashi,
  Phys.\ Rev.\  D {\bf 63}, 035008 (2001)
  [arXiv:hep-ph/0007202].

\bibitem{Ellis:2002wv}
  J.~R.~Ellis, K.~A.~Olive and Y.~Santoso,
  Phys.\ Lett.\  B {\bf 539}, 107 (2002)
  [arXiv:hep-ph/0204192].
\bibitem{Ellis:2002iu}
  J.~R.~Ellis, T.~Falk, K.~A.~Olive and Y.~Santoso,
  Nucl.\ Phys.\  B {\bf 652}, 259 (2003)
  [arXiv:hep-ph/0210205].

\bibitem{Paige:2003mg}
  F.~E.~Paige, S.~D.~Protopopescu, H.~Baer and X.~Tata,
  arXiv:hep-ph/0312045.

\bibitem{isawig}
http://www.hep.phy.cam.ac.uk/~richardn/HERWIG/ISAWIG/

\end{thebibliography}
\end{document}